\documentclass[a4,twocolumn,aip,chaos,showpacs,preprintnumbers,amsmath,amssymb]{revtex4}
\usepackage{epsfig}
\usepackage{graphics}
\usepackage{lipsum}
\usepackage{sidecap}
\usepackage{graphicx} 
\usepackage{dcolumn} 
\usepackage{bm}
\usepackage{sidecap}
\begin{document}
\title{Applicability of 0-1 Test for Strange Nonchaotic Attractors}

\author{R.~Gopal$^{1,2}$}
\author{A.~Venkatesan$^{1}$}
\author{M.~Lakshmanan$^2$}%
 
\affiliation{
$^1$Department of Physics, Nehru Memorial College(Autonomous),Puthanampatti, Tiruchirapalli 621 007, India.\\
$^2$Centre for Nonlinear Dynamics, School of Physics, Bharathidasan University, Tiruchirapalli-620024, India\\
}
\date{\today}
    
\begin{abstract}
 \hspace{1cm}We show that the recently introduced 0-1 test can successfully distinguish between strange nonchaotic attractors(SNAs) and periodic/quasiperiodic/chaotic attractors, by suitably choosing the arbitrary parameter associated with the translation variables in terms of the golden mean number which avoids resonance with the quasiperiodic force. We further characterize the transition from quasiperiodic to chaotic motion via SNAs in terms of the 0-1 test. We demonstrate that the test helps to detect different dynamical transitions to SNAs from quasiperiodic attractor or the transitions from SNAs to chaos. We illustrate the performance of the 0-1 test in detecting transitions to SNAs in quasiperiodically forced logistic map, cubic map, and Duffing oscillator.
\end{abstract}
\pacs{05.45.-a,05.45.Tp, 05.45.Pq}
\maketitle
\textbf{ A strange nonchaotic attractor(SNA) is considered as a complicated structure in phase space. Such a complex structure is a typical property usually associated with a chaotic attractor. However, SNAs are nonchaotic in a dynamical sense because they do not exhibit sensitive dependence on initial conditions (as evidenced by negative Lyapunov exponents). These attractors are  ubiquitous in different quasiperiodically driven nonlinear systems. While the existence of SNAs has been firmly established, a question of intense interest is how do these attractors morphologically differ from other attractors and how they can be characterized and quantified. Two specific characterizations, namely finite time Lyapunov exponents and recurrence plots were proposed earlier in the literature. However, in the present work we show that the recently introduced measure of 0-1 test clearly distinguishes quasiperiodic motion, SNAs and chaotic attractors and also allows one to detect different dynamical transitions from quasiperiodic motion to SNAs and SNAs to chaotic attractors.} 
\section{\label{sec:level1}INTRODUCTION}
\hspace{1cm}A central problem in the study of deterministic dynamical systems is to identify different types of asymptotic behavior of a given system and to understand how the behavior changes as the system parameter changes~\cite{grassberger1993}. The asymptotic behavior can be, for instance, a periodic oscillation, a quasiperiodic motion, a SNA, or a random or a chaotic motion~\cite{ottb1994,lakshmanan2003}. There has been considerable work in the past addressing how dynamical systems develop chaos from periodic or quasiperiodic motions and how they can be characterized by  various diagnostic tools such as  Lyapunov exponents, fractal dimemsion and recurrence analysis. Some among  these like the Lyapunov exponents and fractal dimension also require  substantial amounts of data, free of noise, in order to perform well. Recently, a much simpler test for the presence of deterministic chaos was proposed by Gottwald and Melbourne~\cite{gottwald2004,gottwald2005,gottwald2009}. Their 0-1 test for chaos takes as input a time series of measurements, and returns a single scalar value either 0 for periodic attractors or 1 for chaotic attractors. The test does not require phase space recontruction. The dimension of the dynamical system and the form of the underlying equations are in general irrelevant, and the input is just the time series data. The test returns a unit value in the presence of deterministic chaos and zero otherwise. Recently, many works have been carried out in this connection, including reliability of the 0-1 test ~\cite{hu2005}, characterization of noisy symbolic time series~\cite{christopher2011} and efficient time series detection in FPU lattices~\cite{romera2011}. However all these studies focused only on distinguishing between regular and chaotic motions. In the present work we investigate and characterize the performance of the 0-1 test when applied  to quasiperiodically forced dynamical systems exhibiting SNAs.\\
\hspace*{1cm}SNAs which are commonly found in quasiperiodically forced systems were first described by Grebogi et al~\cite{grebogi1984}. These attractors are geometrically strange as they are properly described by a fractal dimension but the largest nontrivial Lyapunov exponent  is negative, implying nonchaotic dynamics. Briefly speaking, a SNA has a complicated (fractal) structure but does not have a positive Lyapunov exponent. Trajectories separate at rates that are slower than an exponential divergence~\cite{kuznetsov1995}. SNAs have been reported in many physical systems such as the quasiperiodically forced pendulum~\cite{romeiras1987}, biological oscillators~\cite{ding1989}, and quantum particles in quasiperiodic potentials~\cite{bondeson1985}, which are also related to the  Anderson localization in the Schr$\ddot{o}$dinger equation~\cite{ketoja1997}. Also, these exotic attractors were confirmed by an experiment consisting of a quasiperiodically forced, buckled, magnetoelastic ribbon ~\cite{ditto1990}, in analog simulations of a multistable potential~\cite{zhou1992}, neon glow discharge experiment~\cite{ding1997}, chemical oscillations~\cite{ruiz2007} and various electronic circuits~\cite{thamilmaran2006,yang1997}. SNAs can be quantitatively characterized by a variety of methods, including the estimation of Lyapunov exponents and fractal dimension~\cite{kuznetsov1995,prasad1998,venkatesan2000} and spectral properties~\cite{heagy1994,ott1994}. The geometric strangeness of the attractor can be measured through indices such as the phase-sensitivity exponent~\cite{kaneko1984,nishikawa1996,venkatesan1998}, while the chaoticity properties can be studied by examining the  finite-time Lyapunov exponents~\cite{prasad1998,venkatesan1999}. The transition from quasiperiodic to chaotic motion via SNAs can be studied by both the variance of finite time Lyapunov exponents and the recurrence plots~\cite{ngamga2007,venkatesan2001}. The quasiperiodic (torus) attractor  and SNAs are characterized by the Lyapunov exponents with zero and  negative values, respectively. However, for quasiperiodically forced dynamical systems, the values of the Lyapunov exponents are negative for both quasiperiodic attractors(tori) and SNAs. In this case the Lyapunov exponents fail to detect transitions from quasiperiodic dynamics to SNAs. Moreover, analytical methods such as the approximation of quasiperiodic forcing by a series of rational approximations, or the analysis of the phase sensitivity of the dynamics demand a knowledge of the underlying map or differential equations which is often not available. However in the present work for the first time it is reported that the 0-1 test clearly distinguishes tori, SNAs and chaos, from the time series alone. In particular this test indicates a value '0' for torus, a value in between '0' and '1' for SNA, and for chaos a value that tends to '1'.\\
\hspace*{1cm}The mechanisms and routes for the birth of SNAs, their characterization in different dynamical systems, and tools for distinguishing different dynamical transitions to SNAs were reported in ~\cite{feudel2006,prasad2001,kim2003,lai1996,yalcinkaya1998,wang2004,khavanov2000,feudel1995,pikovsky1995,negi2000}. Among the different routes to SNAs, Heagy-Hammel, fractalization and intermittency transitions are ubiquitous in quasiperiodically driven nonlinear systems. In the present paper, we employ the 0-1 test to characterize  different dynamical transitions to SNAs, in particular the Heagy-Hammel, fractalization and intermittent transitions. Specifically, the 0-1 test characterization for transition to SNAs is examined in quasiperiodically driven logistic map, cubic map and  parametrically driven Duffing oscillator to validate our conclusions. \\
\hspace*{1cm}The organization of the paper is as follows. In Sec.II we briefly review the 0-1 test. In Sec.III we report the applicability of the 0-1 test to quasiperiodically driven systems.  Sec.IV examines the characterization using the 0-1 test for different routes of  creation of SNAs, while in Sec.V we discuss the transition from SNAs to chaos. The final section gives our conclusions. In Appendix A we briefly describe our procedure to identify an optimal value for the parameter $c$ required for the 0-1 test. In Appendix B we briefly analyze the behaviors of the translation variables $p$ and $q$, and the asymptotic growth rate $K$ as a function of the length of the time series $N$. In Appendix C we present the nature of attractors in the quasiperiodically forced cubic map using the 0-1 test. Finally, in Appendix D we analyze the distribution of local asymptotic growth rate $P(K,N)$ along  with finite time Lyapunov exponents(FTLEs) for a statistical evaluation of the results. \\
\begin{figure}
\centering
\includegraphics[width=0.8 \columnwidth]{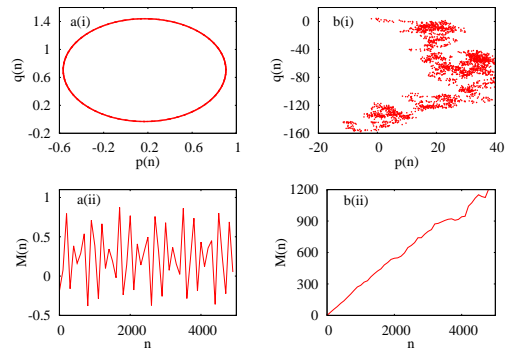}
\caption{\label{fig1}(Color online)Plots of (i) translation variable $p$ versus $q$ (ii) mean square displacement versus $n$ for the logistic map (Eq.(11)) corresponding to (a) periodic dynamics for $\epsilon=0.0$, $\alpha=3.5$ and (b) chaotic dynamics for $\epsilon=0.0$, $\alpha=4.00$. Here, the total length of the time series $N=5\times10^{4}$ with $n=N/10$.}
\end{figure}
\begin{figure}
\centering
\includegraphics[width=0.9 \columnwidth]{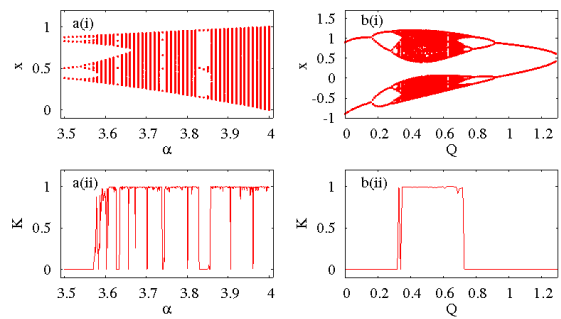}
\caption{\label{fig2}(Color online) Application of 0-1 test for the force-free case of (a) logistic map (Eq.(11)) and (b) cubic map (Eq.(12)): (i) Bifurcation diagram, (ii) Asymptotic growth rate($K$) from 0-1 test as a function of system parameter. Here $N=5\times10^{4}$.}
\end{figure}
\section{UNDERSTANDING OF 0-1 TEST}
 Consider an n-dimensional  dynamical system  $\dot{X}=F(X)$, $X=(x_{1},x_{2},...,x_{n})^{T}$ and denote a solution of the underlying system as $X(t)$. Let $\phi(x)$ be an observable of the underlying system. The 0-1 test uses the observable to drive the dynamics on a well chosen Euclidean extension, and then exploits a theorem from Nicol et al~\cite{nicol2001,ashwin2001}, which states that the dynamics on the group extension is bounded if the underlying dynamics is nonchaotic, but it behaves like a Brownian motion if the dynamics is chaotic. The 0-1 test for chaos first appeared in~\cite{gottwald2004}. It is designed to distinguish chaotic behavior from regular behavior in deterministic systems. The test results in a value 1 if the system is chaotic and a 0 if the system is regular. Theoretical justification was given to the 0-1 test in~\cite{gottwald2009}. An implementation technique was also proposed for the 0-1 test in~\cite{gottwald2009}, which has been very useful for the present work. Here we present a brief outline of the 0-1 test followed by a modified version of the 0-1 test illustrated in ~\cite{gottwald2009,christopher2011}.
\subsection{Translation variables and the 0-1 test for regular and chaotic motion} 
Consider a time series of length $N$. The 0-1 test begins with a computation of two variables $\ p(n)$ and $\ q(n)$ which are the so-called translation variables: 
\begin{equation}
p(n)=\sum_{j=1}^{n} \phi(j) cos(jc),    
\end{equation}
\begin{equation}
q(n)=\sum_{j=1}^{n}\\\phi(j) sin(jc),    
\end{equation}
where\hspace*{0.3cm} $n=1,2,...,N.$\\

Here $\phi(j)$ is an observable constructed from the time series. The value of $c$ may be chosen randomly  in the interval (0,2$\pi$) for nonquasiperiodic systems~\cite{gottwald2009}.  One may note that it is the behavior of these translation variables that determines whether the dynamics of the system is regular or chaotic. The translation variables also exhibit a Brownian motion as a function of $n$, when the dynamics is chaotic. For Brownian motion we expect the average of [$p(j+n)$-$p(j)$] and [$q(j+n)$-$ q(j)$] to grow like $\sqrt{n}$ for $j=1,2,3,4...$ Therefore the square of [$p(j+n)$-$p(j)$] and [$q(j+n)$-$q(j)$] should asymptotically approach a linear growth proportional to $n$ for sufficiently large values. Hence, to find the behavior of the translation variables the 0-1 test calculates their mean square displacement $(D(n))$, 

\begin{equation}
D(n)=\lim_{N \rightarrow \infty} \frac{1}{N} \sum_{j=1}^{N} \left[ p(j+n) - p(j) \right]^{2}
+\left[ q(j+n) - q(j) \right]^{2}
\end{equation}
However, in order to obtain a better convergence of mean square displacement, for numerical purpose,  we have used a modified mean square displacement $M(n)$(which exhibits the same asymptotic growth rate as $D(n)$) of the form as suggested in~\cite{gottwald2009}
 \begin{equation}
M(n)=D(n)-V_{osc}(c,n),
\end{equation}

\noindent where 
\begin{equation}
V_{osc}(c,n)=\lim_{N \rightarrow \infty} \frac{1}{N}  \sum_{j=1}^{N}\phi(j)^{2}. \frac{(1-cos(nc))}{(1-cos c)} . \nonumber
\end{equation}
The theory behind the 0-1 test shows that if the dynamics of the system is regular, then (4) is bounded in time. However if the dynamics of the system is chaotic then (4) scales linearly in time. For numerical purpose, with the limit in (4), we must use $n<<N$. In practice, the limit  $M(n)$ exists when $n=N/10$, where one expects $M(n)$ to scale  linearly. The dynamics is further  characterized through the computation of the asymptotic growth rate $K$ which can be calculated directly from $M(n)$ by using a linear regression for the log-log plot of $M(n)$ given by the defnition,
\begin{equation}
 K =\lim_{n \rightarrow \infty}\frac{log M(n)}{log(n)}.      
\end{equation}
\hspace*{0.5cm}It has been pointed out in~\cite{gottwald2009} that the linear regression method leads to a distortion for small values of $n$. Hence, it has been recommended~\cite{gottwald2009} to estimate the asymptotic growth rate $K$ by using a correlation method which works better than the linear regression method. In the present work we employ the correlation method for the estimation of $K$. In this procedure, $K$ is the correlation coefficient of the vectors
\begin{subequations}
\begin{eqnarray}
&&\xi=(1,2,....,n)^{T} ,\\
&&\hspace{0.5cm}\mbox{and}    \nonumber \\
&&\Delta=(M(1),M(2),....,M(n))^{T}.
\end{eqnarray}
\end{subequations}

Now one can define the correlation coefficient
\begin{equation}
	K=corr(\xi,\Delta)=\frac{cov(\xi,\Delta)}{\sqrt{var(\xi)var(\Delta)}}\in[-1,1] \label{k-eq}.
\end{equation}

\noindent Here
\begin{equation}
 cov(\xi,\Delta)=\frac{1}{n}\sum_{j=1}^{n}(\xi(j)-\bar{\xi})(\Delta(j)-\bar{\Delta}) ,
 var(\xi)=cov(\xi,\xi),			\nonumber
\end{equation}

\noindent where
\begin{equation}
 \bar{\xi}=\frac{1}{n}\sum_{j=1}^{n}\xi(j).	\nonumber
\end{equation}

Finally a value of $K=0$ indicates a non-chaotic data set while a value of $K=1$  indicates a chaotic data set. We use the above analysis to characterize SNAs from quasiperiodic and chaotic attractors by properly choosing the value of 'c' in Eqs.(1)-(2) for quasiperiodic dynamical systems.\\
\hspace*{0.5cm} The functions $\ p(n)$ and $\ q(n)$ in Eq.(1) can be viewed as a component of the solution to the skew product system,
\begin{equation}
   \theta(n+1)=\theta(n)+c+\phi(n),\hspace*{1cm}
\end{equation}
\begin{equation}
 p(n+1)=p(n)+\phi(n)cos(\theta(n)),
\end{equation}
\begin{equation}
q(n+1)=q(n)+\phi(n)sin(\theta(n)), 
\end{equation}\\
driven by the dynamics of the observable $\phi(n)$. Here ($\theta$,$\ p$,$\ q$) represent the coordinates on the Euclidean group rotation $\theta$ and translation $(p,q)$ in the plane. It was argued that an inspection of the dynamics of the ($p$,$q$) trajectories provides a quick and simple visual test of whether the underlying dynamics is regular or chaotic~\cite{gottwald2004,gottwald2005}.

\subsection{Extending the 0-1 test to quasiperiodically forced systems exhibiting SNAs} 
The above 0-1 test, when applied in the standard way to quasiperiodically forced systems, exhibiting SNAs, does not clearly distinguish SNAs from torus and chaotic attractors. The 0-1 test returns either a value 0 or 1 for the SNAs depending on the route by which they are created, as long as the parameter $c$ in Eqs.((1)-(2)) is chosen as an arbitray number in the interval (0,2$\pi$). In fact, in a recent study Dawes and Freeland~\cite{dawes2008} have attempted to distinguish between quasiperiodic attractors and SNAs with a modification of the 0-1 test by the addition of noise with the mean square displacement. In particular, by choosing the value of $c$ to be equally spaced in the interval (0.1,$\pi/2$), they showed that for a quasiperiodic attractor $K$ takes a value zero and for a SNA it takes a value '1' with large deviations for certain values of $c$. However this method does not distinguish SNAs from chaotic attractors. In the present study, we have chosen the value of $c$ in terms of the golden mean number like $(\sqrt{5}+1)/2$ or $\frac{(2\pi)^{2}}{\omega}$, where $\omega=(\sqrt{5}-1)/2$ to  characterize SNAs in quasiperiodically forced dynamical systems, so as to avoid resonance with the translation
 variables. Details are given in Appendix A. With such a choice, we find that we can  clearly distinguish torus, SNA and chaos as well as different dynamical transitions involving SNAs. These are discussed in detail in the following.\\
\hspace*{0.5cm} In our numerical analysis the value of the asymptotic growth rate $K$ is computed by taking the average of it over a number of segments in the given trajectory. That is, a given trajectory is divided into a number of segments each of length $L$ and, in each of these, the corresponding asymptotic growth rate $K_{i}$ (where $'i'$ denotes the segment) is evaluated. Finally the average is taken to compute the value of $K$. Specifically, we consider a total time series of length $N=3\times10^5$ for both quasiperiodically forced logistic map and cubic map while for quasiperiodically forced double well Duffing oscillator we take $N=5\times10^5$ (for continuous systems the data points are considered after  sampling the time series data). In the total time series $L=10000$ data points is considered as the length of each segment.\\

\hspace*{0.5cm} We also note here that the qualitative nature of the attractors, the trajectories in the translation variables $(p,q)$ space, and the mean square displacement essentially remain unchanged as a function of $N$ (as discussed in Appendix B). So whenever these quantities are displayed in the following, we use $N=5\times10^4$ data points, but for the calculation of $K$ we use a much larger time series as mentioned above.
\begin{figure*}
\centering
\includegraphics[width=1.5\columnwidth=0.00,height=10cm]{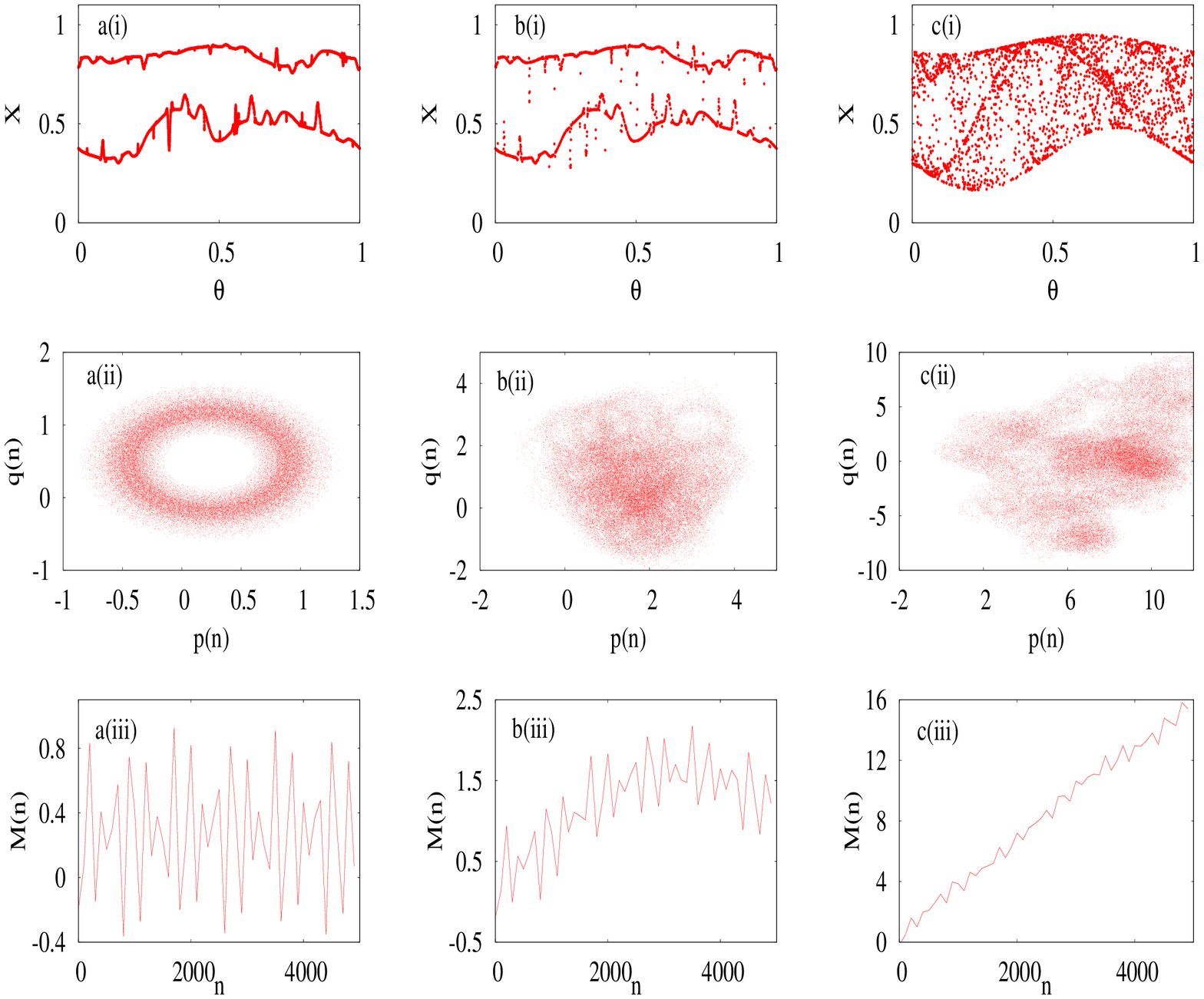}
\caption{\label{fig3}(Color online) For the quasiperiodically forced logistic map (a) Torus motion for $\epsilon'=0.3 $, $\alpha=3.4874$ (b) SNA for $\epsilon'=0.3$, $\alpha=3.488$ and (c) chaotic attractor for $\epsilon'=0.5$, $\alpha=3.6$: (i) Projection of the attractor; (ii) Dynamics of the translation components $(p,q)$ in terms of 0-1 test (Eqs(1)-(2)); (iii) Mean square displacement as a function of $n$. Here $N=5\times10^{4}$ with $n=N/10$.}
\end{figure*} 

\section{Applicability of 0-1 Test in Quasiperiodically driven systems}
In this section we consider the applicability of the 0-1 test for quasiperiodically forced dynamical systems to identify SNAs and distinguish them from periodic, quasiperiodic and chaotic attractors. We choose the following models of quasiperiodically forced dynamical systems for our present study;\\

(i)\textbf{\textit{Quasiperiodically forced logistic map}}:\\
The dynamics is described by the coupled map~\cite{prasad1998},
\begin{subequations}
\begin{eqnarray}
&&x_{n+1}=\alpha[1+\epsilon cos(2\pi\theta_{n})] x_{n}(1-x_{n}), \\
&&\theta_{n+1}=\theta_{n}+\omega\,(mod\, 1),
\end{eqnarray}
\end{subequations}
where $\omega=(\sqrt{5}-1)/2$ is the irrational driving frequency, $\epsilon$ represents the forcing amplitude, and $\alpha$ is a driving parameter. The different dynamical regimes can be  characterized by using the rescaled parameter of the forcing amplitude $\epsilon'=\epsilon/(4/\alpha-1)$ versus $\alpha$.
In this map the transition from torus to SNAs is characterized by the behavior of the largest Lyapunov exponent as well as the characteristic distribution of the finite time Lyapunov exponents. Further, SNAs are created at a saddle node bifurcation, when the dynamics shows type-I intermittency and other transitions.\\

(ii)\textbf{\textit{Quasiperiodically forced cubic map}}:\\ 
The system is given by 

\begin{subequations}
\begin{eqnarray}
&&x_{n+1}=Q+fcos(2\pi\theta_{n})-Ax_{i}+x_{i}^3,  \\ 
&&\theta_{n+1}=\theta_{n}+\omega\,(mod\, 1),
\end{eqnarray}
\end{subequations}
\noindent where $\omega=(\sqrt{5}-1)/2$ is the irrational driving frequency. By varying the values of $A$ and $f$, the creation of SNAs has been shown to occur by means of four distinct routes due to the truncation of torus doubling bifurcations. In particular, the formation of SNAs through Heagy-Hammel, fractalization, type-III intermittency and crises induced intermittency are  described in ~\cite{venkatesan2001}. The SNAs created through different mechanisms are characterized in terms of the Lyapunov exponents, phase sensitivity and finite time Lyapunov exponents.\\

(iii)\textbf{\textit{Damped two frequency parametrically driven double well Duffing oscillator}}:\\
The damped two frequency parametrically driven double well Duffing oscillator is described by the nonlinear differential equation

\begin{equation}
\ddot{x}+H\dot{x}-[1+0.3(R\,cost+cos\,\sigma\,t)]x+x^{3}=0,  \hspace*{0.5cm} (.=\frac{d}{dt})
\end{equation}
which is a well suited model for buckled beam oscillations. Here the driving frequency  $\sigma$ is chosen to be the irrational number $\sigma=(\sqrt{5}+1)/2$. As the parameters $R$ and $H$ are varied one can locate different regions in the parameter space which exhibit different dynamical states. There are several mechanisms through which SNAs are formed in this system, two of which appear to be quasiperiodic analogies of intermittency transition in unforced systems. In addition to these, torus collision route to SNA as well as fractalization route have been identified in ~\cite{venkatesan2000}.

\subsection{Dynamics in the absence of quasiperiodic force}
To start with we  characterize periodic and chaotic attractors in Eqs.(11) and (12) in the absence of  quasiperiodic forcing. For this purpose the  value of $c$ in Eqs.(1) and (2) is randomly chosen in the interval (0,2$\pi$), and the observable is $\phi(j)=x_{j}$. First, the periodic and chaotic behaviors are investigated in the logistic map by using the new coordinates of translation variables $ p(n)$ and $ q(n)$. To analyze the behavior of the translation variables, let us fix the parameters as  $\alpha=3.5$ ($\epsilon=0.0$) in (11). Then the system admits regular dynamics. For  $\alpha=4.00$          ($\epsilon=0.0$) the system exhibits chaotic dynamics. For regular/chaotic dynamics the trajectories in the ($p(n)$, $q(n)$)  space indicate bounded/unbounded (Brownian) motion [see Fig. 1a(i), Fig. 1b(i)], respectively. The bounded and unbounded behaviors in the ($p(n)$, $q(n)$)  space is further investigated in terms of the mean square displacement.  Figs. 1a(ii) and b(ii) indicate mean square displacement versus $n$ which is oscillatory for periodic dynamics, and scales linearly with time for chaotic dynamics. A similar behavior has also  been confirmed for the cubic map with $f=0$ in Eq.(12). \\

\hspace*{0.5cm}Next, we consider the full dynamics of the logistic map given by Eq.(11) and cubic map (Eq.12) in the absence of quasiperiodic forcing. While the logistic map exhibits period-doubling bifurcations as shown in Fig. 2a(i), the cubic map exhibits period doubling route to chaos followed by period halving phenomenon as a function of the system parameter $Q$ which is shown in Fig. 2b(i). In order to confirm the applicability of the 0-1 test, we evaluate the asymptotic growth rate $K$ of the mean square displacement given by  Eq.(7) in Figs. 2a(ii) and b(ii), respectively, for the logistic map and cubic map in the absence of quasiperiodic forcing. As expected the 0-1 test returns a value 0 for periodic motion and 1 for chaotic motion. A similar analysis has been performed for the Duffing oscillator (13) with single periodic force, but we do not present the results here for brevity.

\begin{figure}
\centering
\includegraphics[width=0.7\columnwidth]{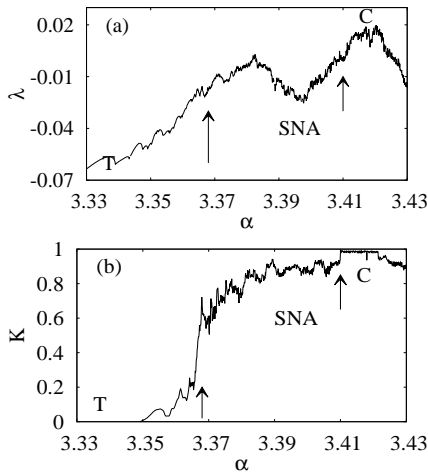}
\caption{\label{fig4}(a) Variation of the largest Lyapunov exponent (b) variation of the asymptotic growth rate ($K$)  from 0-1 test as a function of $\alpha$ for fixed $\epsilon=0.05$ indicating torus$\rightarrow$SNA$\rightarrow$chaos transitions in the quasiperiodically forced logistic map. The vertical arrows in this and the following figures indicate transition points. Here, $N=3\times10^{5}$ with $L=10^{4}$  as the length of each segment.}
\end{figure}
\subsection{Dynamics with quasiperiodic force}
For quasiperiodically  forced  discrete driven systems as well as flows, the dynamics can be  chaotic, nonchaotic or strange nonchaotic depending on the values of the system parameters. Previous studies show that in quasiperiodically driven dynamical systems the regions of SNAs can be quite complicated and also the boundaries between chaotic attractors and SNAs are convoluted, with regions of torus  attractors interspersed among the regions of SNAs~\cite{prasad1998,feudel2006}. We now wish to characterize  SNAs in these quasiperiodically forced  systems by using the 0-1 test to distinguish SNAs with  quasiperiodicity and chaos. Again, we take the time series data (observable) as $\phi(j)=x_{j}$ and the data always contains a component of the quasiperiodic forcing.\\

\hspace*{0.5cm} In the present study, as noted earlier we fix a single value for $c$ in Eqs.(1) and (2) in terms  of an irrational number like $(\sqrt{5}+1)/2$ or $\frac{(2\pi)^2}{\omega}$, where $\omega=(\sqrt{5}-1)/2$ to  characterize SNAs in quasiperiodically forced dynamical systems. We zero down on a most suitable value of $c$ by working on the attractors in the space of the translational variables for a set of values of $c$ in terms of the golden mean for typical cases of torus, SNAs and chaos in the quasiperiodically forced systems (11)-(13). We pick up that value of $c$ which most closely replicates the expected nature of the attractors: torus like structure for quasiperiodic attractors, bounded  motion for SNAs and unbounded motion for chaotic attractors. The details are given in Appendix A. By this procedure we find that a choice $c=\frac{(2\pi)^2}{\omega}$, where $\omega=(\sqrt{5}-1)/2$, is the most suitable value for the present study.\\

\hspace*{0.5cm} The different possible dynamical behaviors are characterized in the logistic map, by varying the values of  $\epsilon'=\epsilon/(4/\alpha-1)$ and  $\alpha$. For instance, for $\epsilon^{'}=0.3$, $\alpha=3.4874$ the system admits torus like motion; for $\epsilon'= 0.3$, $\alpha = 3.488$ it is a SNA, while for $\epsilon'= 0.5 $, $\alpha = 3.6$, it is a chaotic attractor~\cite{prasad1998}. Figures \ref{fig3} a(i),b(i), and c(i) clearly indicate the projection of the dynamical attractors in the ($x$, $\theta$) space. Then, the dynamics of translation variables in the ($p(n)$,$q(n)$) space indicates that the points are uniformly distributed over a circle for quasiperiodic (torus) regime (Fig.3a(ii),  Brownian like behavior for chaotic regime (Fig.3c(ii)), and for SNAs the ($p(n)$,$q(n)$) space shows that the points are uniformly distributed and bounded with minimal Brownian like behavior (Fig.3b(ii)). We characterize the behavior of the translation variables by using the mean square displacement$(M(n))$ as a function of $n$, and find that $M(n)$ is oscillatory for quasiperiodic regimes (Fig.3a(iii)); grows linearly with time for chaotic attractors (Fig.3c(iii)), and for SNAs $M(n)$ grows linearly initially with oscillatory motion, but tends to get bounded for large $n$ (Fig.3b(iii)).  The aforementioned analysis clearly shows that the 0-1 test allows one to  distinguish SNAs from quasiperiodic regime and chaos. A similar analysis has also distinguished tori, SNAs and chaotic attractors in the quasiperiodically forced cubic map represented by Eq.(12). For details, see Appendix C.\\
\hspace*{0.5cm} We next study the  characterization of  different dynamical attractors as a function of the parameter $\alpha$ in the quasiperiodically forced logistic map (Eq.11) in the parameter regime $\alpha$ $\in(3.33,3.41)$  by using the 0-1 test.  The logistic map (Eq.11) has been already investigated and various transitions to SNAs have been identified using tools such as Lyapunov exponents and their variance, finite time Lyapunov exponents and power spectral measures~\cite{prasad1998,venkatesan2001}. Here, we consider the above parameter regime investigated in Ref.~\cite{prasad1998} for our 0-1 test analysis also. To compare our results using the 0-1 test with the already reported results on torus
$\rightarrow$SNA$\rightarrow$ chaos transitions, here we present the nature of the largest Lyapunov exponent along with the asymptotic growth rate $K$ using the 0-1 test as shown in Fig.4. As noted above our present study of the 0-1 test using the value of $c$ = $\frac{(2\pi)^{2}}{\omega}$, where $\omega=(\sqrt{5}-1)/2$, indicates that the asymptotic growth rate returns a value 0 for torus regime, 1 for chaotic regime, while in the SNA regime the value of $K$ varies in between 0 and 1 non-monotonically. It may be also noted that there is an  abrupt transition in the value of $K$ from a small value to a larger one (indicated by a vertical arrow below $\alpha=3.37$) during the transition from torus to SNA. Between the value of $\alpha$, 3.35$\leqslant\alpha\leqslant$3.37, the torus gets wrinkled before transition to a SNA which is indicated  by a nonzero value of $K$. One also notes that while the Lyapunov exponent ($\lambda$) as a function of $\alpha$ (Fig.4(a)) does not clearly distinguish between torus and SNA, where it takes negative value in both the regimes, the asymptotic growth rate $K$ clearly distinguishes the torus and SNA regimes, as well as the chaotic regime.\\ 
\hspace*{0.5cm} Similar analysis has been carried out for the quasiperiodically forced cubic map and Duffing oscillator to confirm that with the choice of $c$ = $\frac{(2\pi)^{2}}{\omega}$, where $\omega=(\sqrt{5}-1)/2$, as an irrational number, the asymptotic growth rate $K$ varies between 0 and 1.

\section{Different Routes to SNAs}
One of our main concerns in this work is whether the 0-1 test can  distinguish different dynamical transition  scenarios involving SNAs through appropriate characterizations. In this section we show clearly that the formation of SNAs through different routes exhibits distinctly clear signatures in terms of changes in the growth rate of $K$. In particular we demonstrate this for the  ubiquitous routes, namely Heagy-Hammel route on torus doubling to chaos via SNAs~\cite{heagy1994}, fractalization of torus leading to SNAs~\cite{kaneko1984,nishikawa1996}, and the appearance of SNAs through type -I intermittent phenomenon~\cite{prasad1998}. For other routes also, one can identify similar signatures. We also note here that each one of the routes has specific significance as discussed in the literature~\cite{ngamga2007,venkatesan2001,feudel2006,prasad2001}: (i) Heagy-Hammel (HH) route: Collision of doubled torus with its unstable parent leading to SNA before the onset of chaos. (ii) Fractalization route: Gratual fractalization of torus giving rise to SNA before the onset of chaos. (iii) Intermittency route: Analog of saddle node bifurcation leading to the birth of SNA before the onset of chaos.

\subsection{Heagy-Hammel Route}
The first route we consider is the Heagy-Hammel(HH) route. In this route, a SNA  occurs when the period doubled torus collides with its unstable parent. That is, the period-$2^{n}$ torus gets wrinkled and upon collision with its unstable parent, a period-$2^{n-1}$ band SNA is formed. To exemplify the nature of this transition, let us consider the quasiperiodically forced logistic map (Eq. 11), where we fix the parameter $\epsilon'=0.3$, while $\alpha$ is varied. In this case, the HH transition takes place at $\alpha_{hh}\simeq3.487793$~\cite{prasad1998}. When we examine the neighborhood of $\alpha_{hh}$, the transition from torus to SNAs is characterized by the Lyapunov exponent($\lambda$) as shown in Fig. 5a(i). Particularly in the torus region,  $\alpha\leqslant \alpha_{hh}$, $\lambda$ varies smoothly whereas in the SNA region, $\alpha\geqslant \alpha_{hh}$, the variation in $\lambda$ is irregular. To validate the applicability of the 0-1 test for SNAs, we consider the aforementioned parameter region and analyze the asymptotic growth rate $K$ versus $\alpha$. One can see that in Fig. 5a(ii) $K$ becomes minimal, that is  '0' or nearly '0', in the torus region. When $\alpha\geqslant \alpha_{hh}$, the growth rate $K$ is varying between 0.2 and 0.7 non-monotonically. The behavior of $K$ versus $\alpha$ shows a transition from regular dynamical regime to a strange nonchaotic one with a threshold value $\alpha_{hh}$, separating both regimes.\\

\hspace*{0.5cm}Similarly, in the quasiperiodically forced cubic map~\cite{venkatesan2001} given by Eq. (12), we fix the system parameter at $f=0.7$ and vary $A$ in the range 1.8865 to 1.8875. Initially for $A\simeq1.8865$ the system has a period doubled torus. Then as $A$ increases to the value 1.8868, the torus begins to wrinkle and approaches its unstable parent. After this the torus collides with the parent and gives birth to a fractal attractor (SNA), when $A$ reaches the value 1.88697. For this value of $A$, the attractor possesses a geometrically strange and nonchaotic behavior~\cite{venkatesan2001}. It is noticed that  these dynamical changes are clearly distinguished in the 0-1 test as shown in  Fig. 5b(ii), while the corresponding Lyapunov exponent structure is shown in Fig. 5b(i). For the torus region the value of $K$ is close to '0', and in the SNA region it oscillates between 0.2 and 0.4. We can again note an abrupt transition in the value of $K$ at the transition point $A=1.88697$ when the torus gets transferred to a SNA.\\

 \hspace*{0.5cm}We conclude that the results of 0-1 test indicate in both the quasiperiodically driven logistic and cubic maps  that the value of  $K$ is smooth in the torus region and, at the critical value, there is a sudden jump after which $K$ oscillates with irregular behavior.\\
\hspace*{0.5cm}The HH  analysis has also been carried out for the parametrically driven Duffing oscillator (Eq. (13)) with the system parameter fixed as $R=0.3$, while the value of $H$ is varied between 0.181 to 0.186. Here one finds that the transition from a SNA to a torus takes place at $H_{c}=0.184$ ~\cite{venkatesan2000}. The Lyapunov exponent $\lambda$ does not show any distinct change in its behavior during this transition. Figure 6(a) is a plot of  $\lambda$ as a function of $H$, in the neighborhood of $H_{c}$, where the variation of $\lambda$ is small and smooth, but in the SNA phase the variation of $\lambda$ is large. On the other hand, the plot between the asymptotic growth rate $K$ and the parameter $H$ indicates as shown in Fig. 6(b) that the value of $K$ sharply (but smoothly) decreases from 0.78 to 0.00 within a narrow range of the parameter $H$ with respect to the transition from SNA to the torus regime, again giving a clear distinction between regions of SNAs and quasiperiodic attractors. We may also note that the HH transition is quite narrow in the parameter range $H$ as seen in Fig.6.

\begin{SCfigure*}
\centering
\includegraphics[width=1.5\columnwidth=0.9]{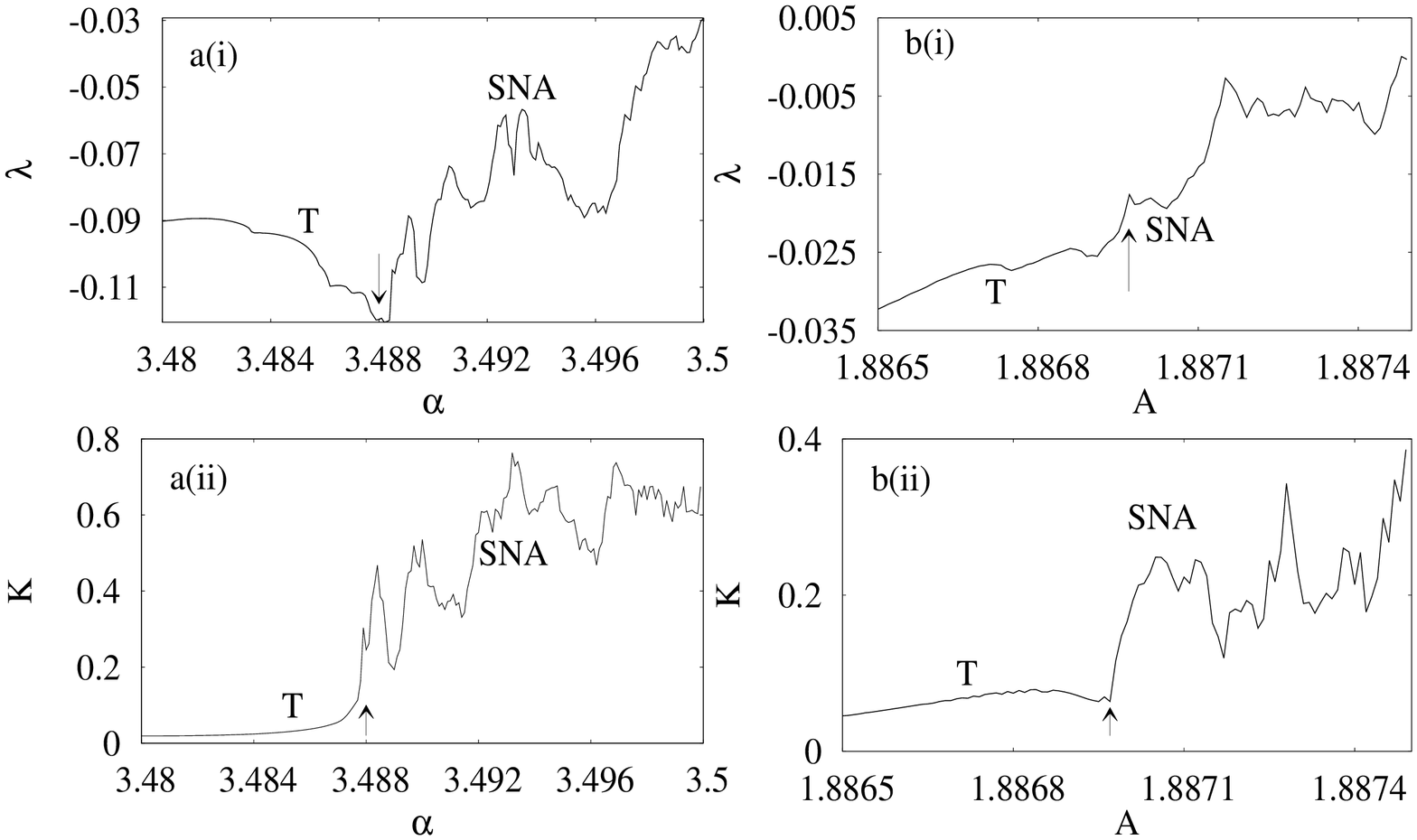}
\caption{\label{fig5}Transition from torus to SNA through HH mechanism in the (a) logistic map (b) cubic map:(i) Behavior of the Lyapunov exponent($\lambda$) (ii) Behavior of the asymptotic growth rate ($K$) from 0-1 test. Here, $N=3\times10^{5}$ with $L=10^{4}$.}
\end{SCfigure*}
\begin{figure}
\centering
\includegraphics[width=0.800\columnwidth]{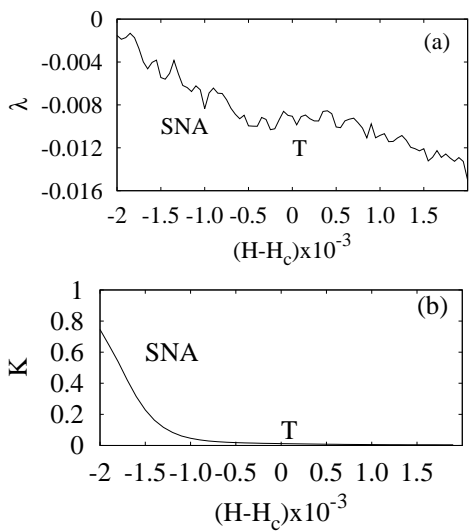}
\caption{\label{fig6}Transition from SNA to torus through HH mechanism in the Duffing oscillator (a) Behavior of the Lyapunov exponent($\lambda$) (b) Behavior of the asymptotic growth rate ($K$) from 0-1 test. Here, $N=5\times10^{5}$ (after sampling the data points) and $L=10^{4}$.}
\end{figure}

\subsection{Fractalization Route}
\begin{SCfigure*}
\centering
\includegraphics[width=1.5\columnwidth=1.00]{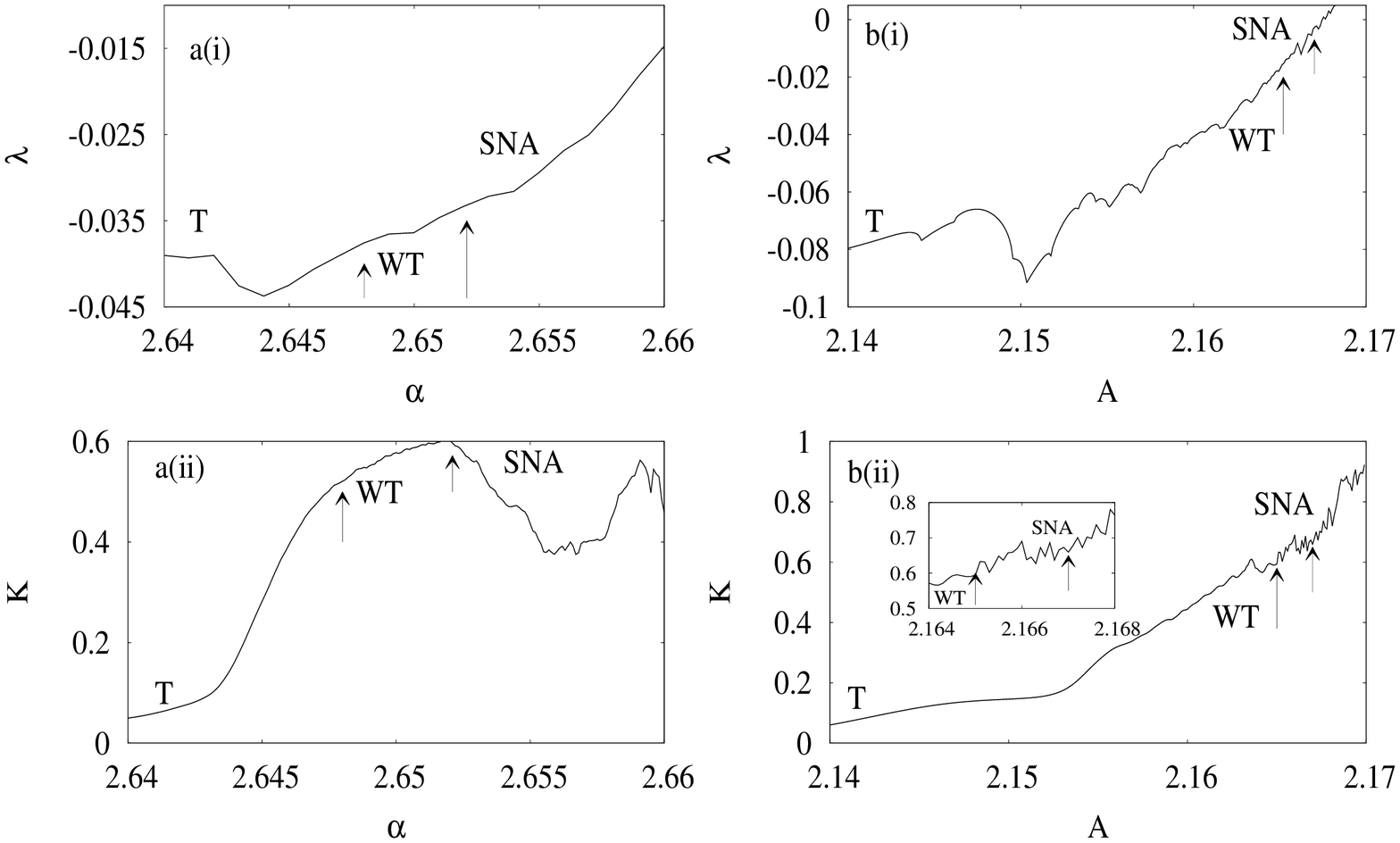}
\caption{\label{fig7}Transition from torus to SNA during fractalization route in the (a) logistic map (b) cubic map: (i) Behavior of Lyapunov exponent($\lambda$) and (ii) Behavior of asymptotic growth rate ($K$) from 0-1 test. Here, $N=3\times10^{5}$ with $L=10^{4}$ as the length of each segment.}
\end{SCfigure*}

 The fractalization route is a common and rapidly changing intriguing transition to SNAs as pointed out by  Nishikawa and Kaneko ~\cite{kaneko1984}. Here a period $n$ torus attractor gets wrinkled and eventually forms a $n$-band SNA. Now we examine the fractalization route in the quasiperiodically forced logistic map (Eq. (11)), where we fix $\epsilon'=1.00$ and the value of $\alpha$ is varied from 2.64 to 2.66. Here, the variation of $\lambda$ as a function of the control parameter $\alpha$ as shown in Fig. 7a(i) describes the transition from a torus to a SNA (through a wrinkled torus (WT)) which occurs at $\alpha_{F}=2.6526$~\cite{prasad1998}. The Lyapunov exponent $\lambda$ is smooth and linear with no particular signature observed during this transition. However, as shown in  Fig.\ref{fig7}a(ii) the results of the 0-1 test indicate that the transition from tori to SNAs via WT is characterized by the  asymptotic growth rate $K$  which grows  linearly from torus to WT, and becomes irregular at the transition for SNA at $\alpha_{F}=2.6526$.\\

\begin{figure}
\centering
\includegraphics[width=0.8\columnwidth]{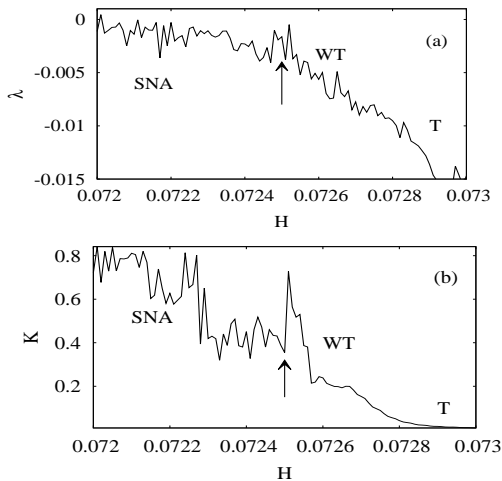}
\caption{\label{fig8}Transition from SNA to torus (a) Behavior of the  Lyapunov exponent($\lambda$) (b) Behavior of the asymptotic growth rate ($K$) from 0-1 test during the fractalization route in the parametrically driven double well Duffing oscillator. Here $5\times10^{5}$ (after sampling the data points) with $L=10^{4}$.}
\end{figure}

\hspace*{0.4cm}To study the fractalization mechanism in the cubic map(Eq.(12)), we fix the parameter at $f=0.1$, and vary $A$ in the range 2.14 to 2.17. For $A=2.14$ the system admits a torus like motion; For $A=2.16$ the system exhibits quasiperiodic oscillations of the doubled  torus and when it is increased to $A=2.165$, the doubled torus becomes a WT, which persists up to $A=2.167$, where the oscillatory structure(wrinkled) gradually approaches a fractal structure(SNA)~\cite{venkatesan2001}. This region is examined by both the Lyapunov exponent($\lambda$) and the 0-1 test. Again the Lyapunov exponent $\lambda$ does not show any specific signature when the transition takes place [See Fig. 7b(i)]. But in the case of the 0-1 test as shown in Fig. 7b(ii) the  value of asymptotic growth rate $K$ is nearly 0 in the torus region and is linearly increasing during transition from torus to WT, and exhibits an irregular behavior at the transition to SNAs.\\

\hspace*{0.4cm} The fractalization route occurs in the parametrically driven Duffing oscillator with $R=0.3$, and $H$ in the range $H={(0.072,0.073)}$. The transition from a SNA to a torus occurs at $H_{F}=0.0729$~\cite{venkatesan2000}. We find that at this transition value of $H$ the asymptotic growth rate $K$ decreases to zero from the SNA region to the torus one [see Fig.8(b)], whose Lyapunov exponent is as shown in Fig.8(a). Again during the transition there is an irregular behavior in $K$ during transition from SNAs to WT, while it is fairly smooth and linear (with small fluctuations) in $K$ from WT to torus. One can conclude that the  asymptotic growth rate $K$ behaves differently in the fractalization route compared to the HH mechanism.
\begin{SCfigure*}
\centering
\includegraphics[width=1.5\columnwidth=1.00]{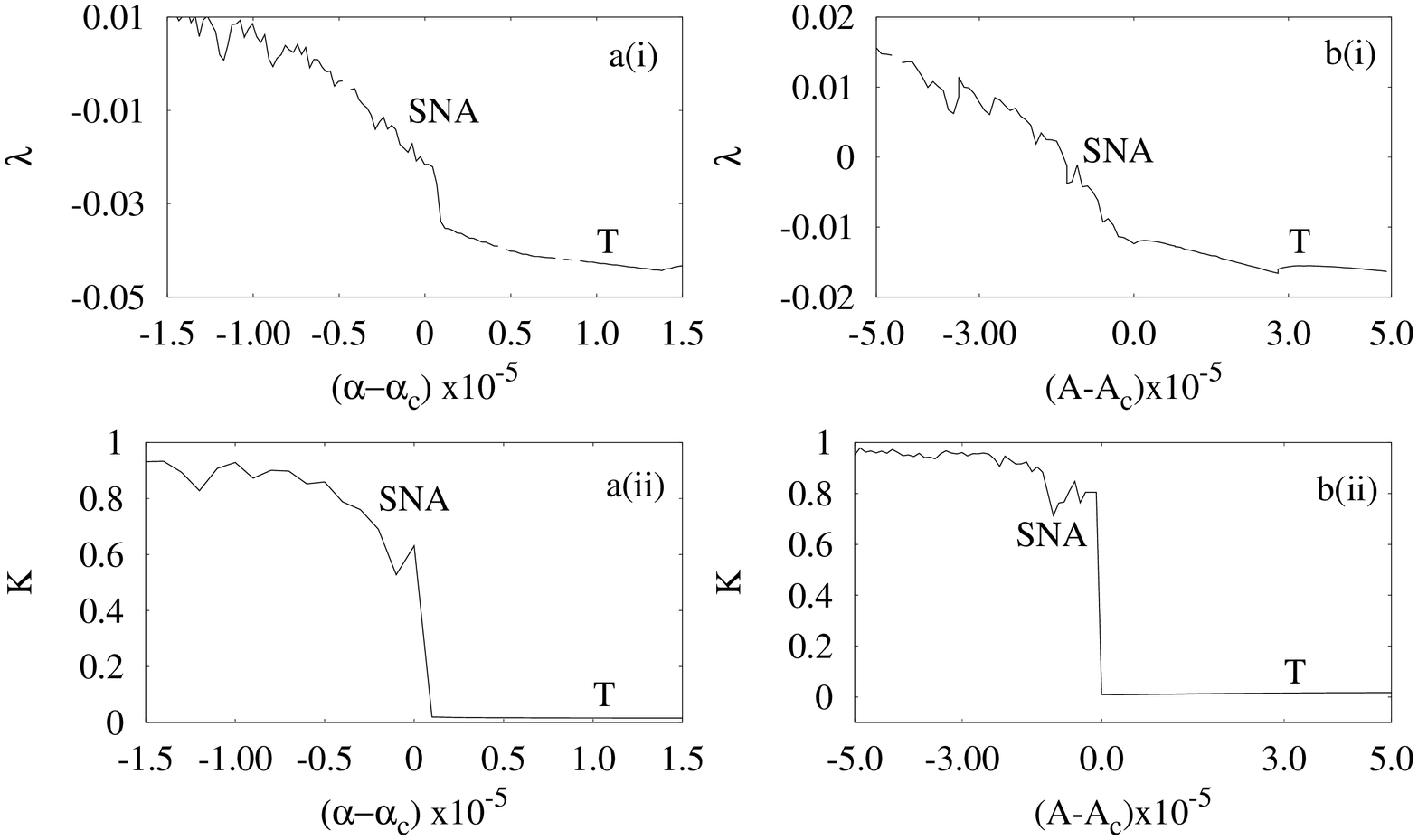}
\caption{\label{fig9}Transition from a SNA to a torus through type I intermittency in the (a) logistic map (b) cubic map:(i) Behavior of the Lyapunov exponent($\lambda$) (ii) Behavior of the asymptotic growth rate ($K$)  from 0-1 test. Here $N=3\times10^{5}$ with $L=10^{4}$.}
\end{SCfigure*}
\begin{figure}
\centering
\includegraphics[width=0.8\columnwidth]{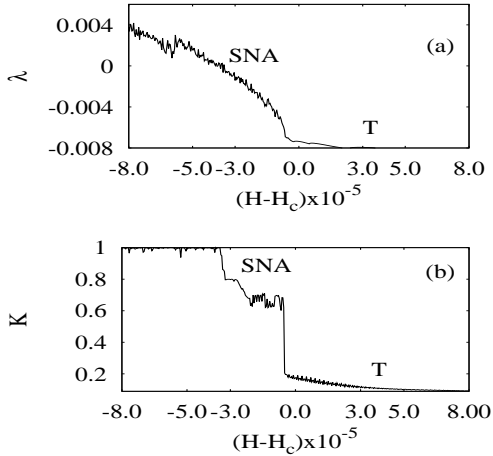}
\caption{\label{fig10}Transition from intermittent SNA to torus motion in the two frequency parametrically driven double well Duffing oscillator:(a) Behavior of the Lyapunov exponent($\lambda$) (b) Behavior of the asymptotic growth rate ($K$)  from 0-1 test. Here, $N=5\times10^{5}$ (after sampling the data points) with $L=10^{4}$ as the length of each segment.}
\end{figure}
\subsection{Intermittency Route}
Finally we discuss the intermittency route for SNA, which differs from those formed through other mechanisms. In ~\cite{prasad1998} the mechanism for the route in the logistic map has been studied. It has been shown that in this route a SNA disappears and is replaced by a one-frequency torus through an analogy with the saddle node bifurcation. In the logistic map [Eq. (11)] the intermittent transition from a SNA to a torus occurs at $\alpha_{c}=3.4058088$ with $\epsilon'=1.00$~\cite{prasad1998}. The intermittent transition from a SNA to a torus in the cubic map(Eq. (12)) is studied in\cite{venkatesan2001}. For the forcing amplitude $f=0.7$, a SNA occurs at $A=1.801685$; On increasing the value of $A$, an intermittent transition occurs at $A_{c}=1.8017$  which indicates a torus motion.\\

\hspace*{0.5cm} In Fig.9(a) we present the nature of the Lyapunov exponent in the neighborhood of the transition point in the system parameter both for the logistic map (Fig.9a(i)) and cubic map (Fig.9b(i)).
 At the intermittent transition, the Lyapunov exponent does not show any significant variation during the transition from SNAs to torus.  On the other hand the 0-1 test  detects intermittent SNAs and torus clearly in Figs. 9a(ii) and b(ii), where the asymptotic growth rate $K$ fluctuates in the intermittent SNA region while it smoothly varies in the torus region. The value of $K$ exhibits a sharp transition from a positive value (nearly 0.8) to zero during the transition from SNAs to torus.\\

\hspace*{0.5cm}Further we consider the intermittent transition in the parametrically driven double well Duffing oscillator [Eq. (13)] with $R=0.35$, while the value of $H$ is varied. For $H=0.19080564$, the system admits a chaotic attractor; a SNA appears at $H=0.190833$. Further  when the value of $H$ is increasing, an intermittent transition from a SNA to a torus occurs at $H_{c} = 0.19088564$~\cite{venkatesan2000}. At this transition point, the Lyapunov exponent [Fig.10(a)] is of decreasing nature, where for  both intermittent SNA and torus the value of $\lambda$ is negative. The 0-1 test clearly detects changes in the growth rate as shown in Fig.10(b).

\begin{figure}
\centering
\includegraphics[width=0.8\columnwidth]{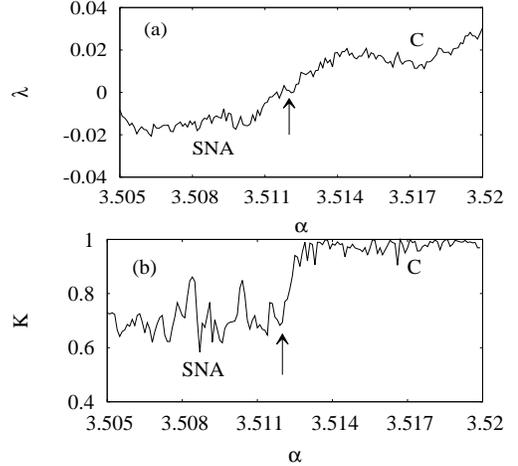}
\caption{\label{fig11}(a) Variation of the Lyapunov exponent $\lambda$ and (b) variation of the asymptotic growth rate ($K$) from 0-1 test as a function of $\alpha$ indicating transition from SNAs to chaotic attractors in the quasiperiodically forced logistic map  with $\epsilon^{'}=0.30$. Here $N=3\times10^{5}$ and $L=10^{4}$.}
\end{figure}
\begin{figure}
\centering
\includegraphics[width=0.8\columnwidth]{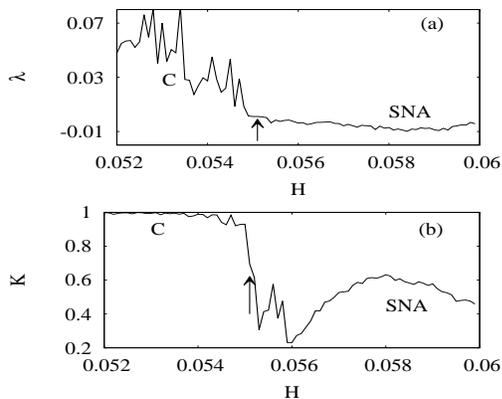}
\caption{\label{fig12}(a) Varition of the Lyapunov exponent $\lambda$ and (b) variation of the asymptotic growth rate $K$ as a function of $H$ indicating transition from chaos to SNAs in the parametrically driven double well Duffing oscillator with R=0.47. Here  $N=5\times10^{5}$ (after sampling the data points) with $L=10^{4}$.}
\end{figure}
\section{The Transitions from SNAs to Chaotic attractors}
In the previous sections we have characterized different routes for the transition between tori and SNAs, which are qualitatively distinguished in terms of the asymptotic growth rate $K$ by the 0-1 test. It has been shown that the value of $K$ shows an abrupt change followed by an irregular behavior in general after the birth of SNAs. In this section we mainly concentrate on important dynamical characteristics of transition from SNAs to chaotic attractors or vice versa, usually characterized by the largest Lyapunov exponent which changes from negative to positive values~\cite{prasad1998,lai1996,yalcinkaya1998}. In general SNAs occur in narrow parameter ranges, and finally get transferred into chaotic attractors. The transition is not usually accompanied by any major change in the form or shape of the attractor. We now study this type of transition in the quasiperiodically forced logistic map [Eq.(11)] with parameter $\epsilon'=0.3$. Negi $et\, al$ ~\cite{negi2000} have found that the transition from SNAs to chaos occurs at $\alpha=3.512$. Fig. 11(a) presents the Lyapunov exponent($\lambda$) for the transition from SNAs to chaos which clearly shows that  $\lambda$  changes from negative to positive values. In the case of the 0-1 test, the value of the asymptotic growth rate $K$ oscillates between 0.5 and 0.8 in the SNA regime,  and suddenly goes to '1' in the chaotic regime  [See Fig.11(b)]. Similarly we have also carried out a study of the transition from  chaos to SNAs in the parametrically driven Duffing oscillator [Eq.(13)] with $R=0.47$ while  varying $H$. For $H=0.05$, the attractor is chaotic. As $H$ is increased to 0.0558, one obtains a SNA. The 0-1 test shows a specific signature of transition from chaos to SNAs  as shown in Fig. 12(b) whose Lyapunov exponent is shown in Fig. 12(a).
\begin{figure*}
\centering
\includegraphics[width=1.50\columnwidth]{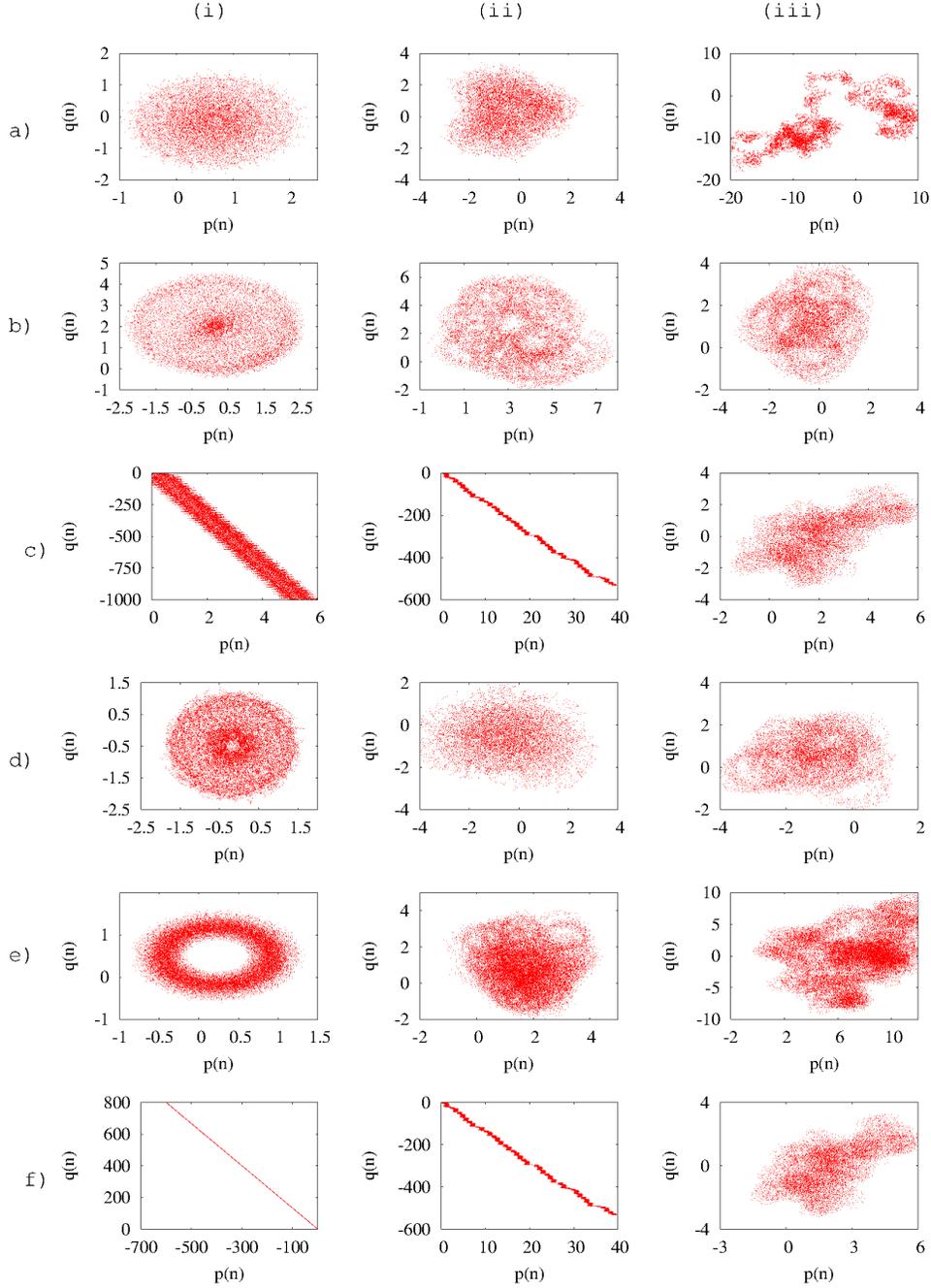}
\caption{\label{fig13}(Color online) Choosing the most optimal value of $c$: Attractors of the quasiperiodically forced logistic map in the translation variables $(p,q)$ space. (i) torus, (ii) SNA and (iii) chaotic attractors for the following choices of $c$: (a)$\frac{\sqrt(5)+1}{2}=1.618034$  (b)$\frac{\sqrt(5)-1}{2}=0.618034$ (c)$\frac{2\pi}{\omega} (\mod 2\pi)= 3.883222$ (d)$\frac{(2\pi)^{3}}{\omega} (\mod 2\pi) =5.513002$ (e) $\frac{(2\pi)^{2}}{\omega} (\mod 2\pi)= 1.045568$ (f) $\frac{2\pi}{\sigma} (\mod 2\pi) =3.883222$, where $\omega=\frac{\sqrt5 -1}{2}$  and $\sigma=\frac{\sqrt 5 +1}{2}$   Note that the choice (e) with $c=\frac{(2\pi)^{2}}{\omega}$,  $\omega=\frac{\sqrt5-1}{2}$ gives the most satisfactory description of all the three attractors. Here $N=5\times10^{4}$.}
\end{figure*}
\section{Conclusion}
In this work we have studied  different routes to SNAs and transitions from tori to SNAs and then to chaotic attractors in  prototypical examples, namely the quasiperiodically forced logistic map, cubic map and parametrically driven Duffing oscillator. In these systems we have studied the applicability of 0-1 test to distinguish SNAs from quasiperiodic motion and chaos, and have shown that the test clearly distinguishes SNAs from other attractors. The present study has also been able to detect the strange nonchaotic dynamics in the Heagy-Hammel, fractalization and intermittecy routes, which are in general not easily detected using the Lyapunov exponent. We have pointed out that in each of the routes, the 0-1 test clearly detects an abrupt change in the asymptotic growth rate $K$ at critical parameter values during transitions.\\
\hspace*{0.5cm}In particular for the HH route the value of $K$ is smooth in the torus region and at a critical value there is a sudden jump after which it exhibits an irregular behavior. In the fractalization route, the value of $K$ grows linearly from torus to WT and becomes irregular at the transition to SNA. For intermittency route the value of $K$ exhibits a sharp transition from a fluctuating value (nearly 0.8) to zero during the transition from SNAs to torus. Finally, our analysis shows that for quasiperiodic attractors the value of $K$ is zero, and if chaotic it is 1, whereas for SNAs it takes values in between '0' and '1'.\\
\hspace*{0.5cm} Thus the tools of 0-1 test helps one to distinguish SNAs from other attractors in a clear manner. Further, we note that the advantages of 0-1 test are that it requires minimal time series data, does not require any equation of motion and is fairly easy to compute. This test has some disadvantages in that it depends upon a properly chosen value of '$c$' in Eq.(2).  In this paper we have used '$c$' in terms of suitable irrational numbers to characterize quasiperiodically forced dynamical systems. Finally, as it requires minimal time series, we believe that this test is more suitable for experimental situations in order to distinguish SNAs from other regular and irregular behaviors.

\begin{acknowledgments}
The work of R.G and M.L. has been supported by the Department of Science 
and Technology (DST), Government of India sponsored IRHPA research project. M.L. is also supported by a Department of Atomic Energy Raja Ramanna fellowship and a DST Ramanna program. A.V. acknowledges the support from UGC minor research project.
\end{acknowledgments}

\begin{appendix}
\section*{Appendix A: Choosing the parameter $c$ of the 0-1 test in quasiperiodically driven systems}
In this appendix we illustrate the method  by which a suitable choice for the parameter $c$ in Eqs.(1)-(2) of the 0-1 test can be made using the quasiperiodically forced logistic map (11) as a specific example. The same analysis has been performed for the quasiperiodically driven cubic map and Duffing oscillator. Here we present the nature of different dynamical attractors in the quasiperiodically forced logistic map, which has been studied in Sec.III.B, and examine the dynamics in terms of the translation variables $(p,q)$ for tori, SNAs and chaotic attractors for different values of $c$ chosen in terms of irrational numbers. The results indicate that while many values of $c$ fail to distinguish quasiperiodic/SNA/chaotic attractors, there exists a set of values which clearly distinguishes them as shown in Figs.13. Particularly, we find that for the choice $c$ = $\frac{(2\pi)^{2}}{\omega}$, where $\omega=(\sqrt{5}-1)/2$, (see Figs.13e) very clear distinction between different attractors can be made. Note that this choice does not produce any resonance with the $(p,q)$ variables. The same value works quite well in the cases of quasiperiodically driven cubic map and Duffing oscillator.  So in our analysis we fix $c$ as $c$ = $\frac{(2\pi)^{2}}{\omega}$, where $\omega=(\sqrt{5}-1)/2$. A second choice is $c=(\sqrt{5}+1)/2$, corresponding to Fig.13(a), which we find to be reasonably satisfactory, though not as good as $c=\frac{(2\pi)^{2}}{\omega}$, where $\omega=(\sqrt{5}-1)/2$. 

\begin{figure}
\centering
\includegraphics[width=1.0\columnwidth=0.6,height=10cm]{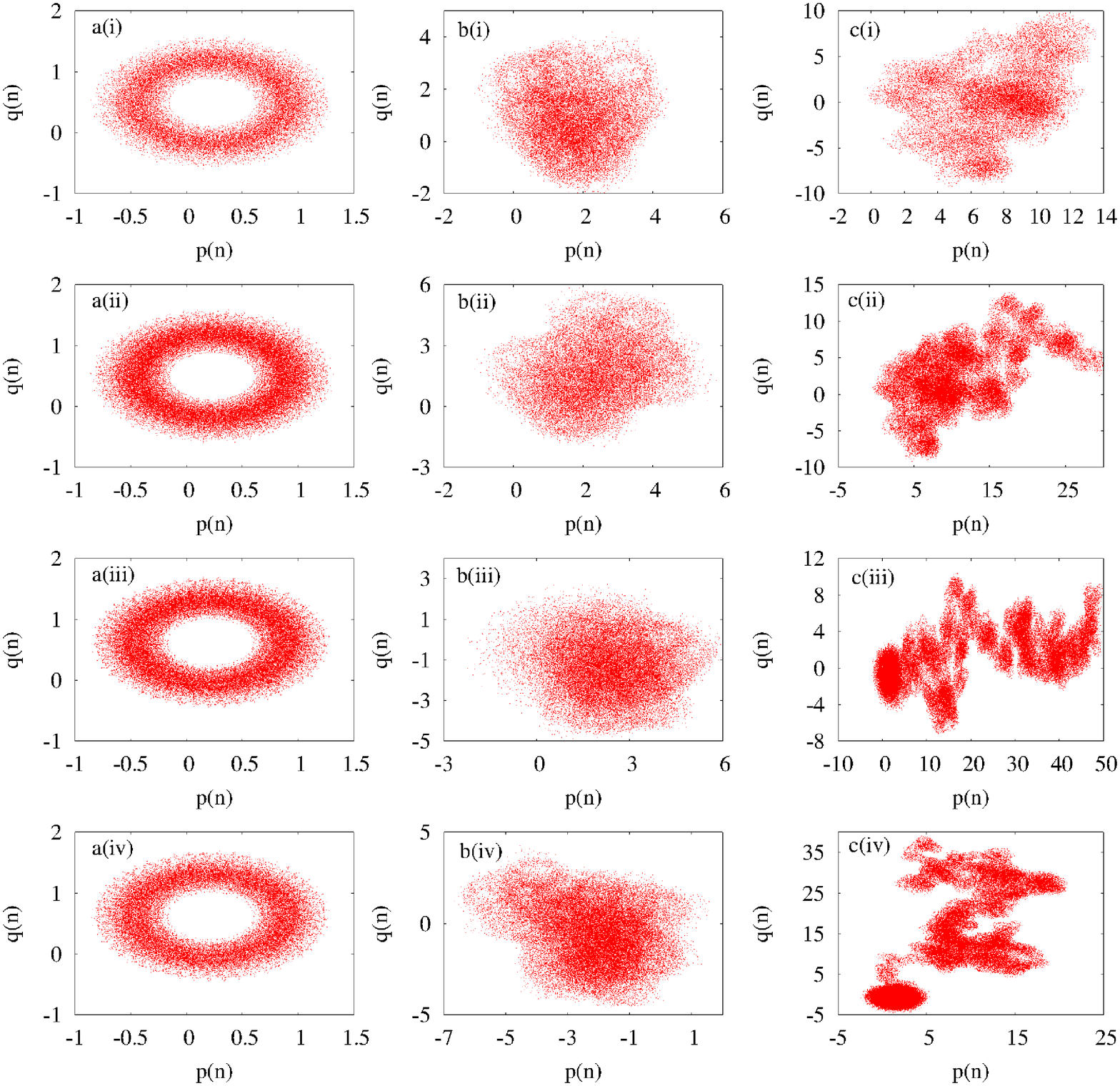}
\caption{\label{fig14}(Color online)Nature of the translation variables $(p,q)$ for the quasiperiodically forced logistic map (Eq.(11)) as a function of the length of the time series $N$ for (a) Torus  for $\epsilon'=0.3 $, $\alpha=3.4874$ (b) SNA for $\epsilon'=0.3$, $\alpha=3.488$ (c) chaotic attractor for $\epsilon'=0.5$, $\alpha=3.6$: (i) $N=5\times10^{4}$ (ii) $N=1\times10^{5}$ (iii) $N=2\times10^{5}$ (iv) $N=3\times10^{5}$.}
\end{figure}

\section*{Appendix B: Behaviors of the translation variables and asymptotic growth rate as a finction of $N$}
In order to investigate how our results behave as a function of the length $N$ of the time series, we discuss the dynamics of the translation variables $(p,q)$ as well as variation of the asymptotic growth rate $K$ as a function of $N$ for the quasiperiodically forced logistic map. We have plotted the variables $(p,q)$ in Figs.14 for different types of attractors (Eq.(11)) for different lengths $N$ of time series data.  From the figure, one can see first that the $(p,q)$ trajectories of the torus motion are bounded for $N=5\times10^{4}$ (Fig.14.a(i)). Then, upon increasing $N$ the $(p,q)$ trajectories continue to be bounded (Figs. 14a(ii)-14.a(iv)) even upto $N=3\times10^{5}$.  Next, we demonstrate in Fig.15 that the asymptotic growth rate $K$ takes a value close to zero irrespective of the length of time series (Fig.15(a)).  We also include the corresponding plot of the Lyapunov exponent in Fig.15(b) for comparision.  We have also carried out similar analysis for the SNAs and chaotic attractors which are reported in Figs. 14(b) and 14(c) respectively.  The corresponding asymptotic growth rate $K$ and Lyapunov exponents are depicted in Figs. 15(a) and (b) respectively (Note that in Fig.15(a), the value of $K$ is shown without taking averages over segments in order to depict its behavior as a function of $N$). The above results confirm that in the cases of SNA and chaotic attractors also the dynamics remains qualitatively unchanged irrespective of the length of the time series. Finally, the results have also been confirmed for the quasiperiodically forced cubic map and Duffing oscillator.

\begin{figure}
\centering
\includegraphics[width=0.8\columnwidth=0.8]{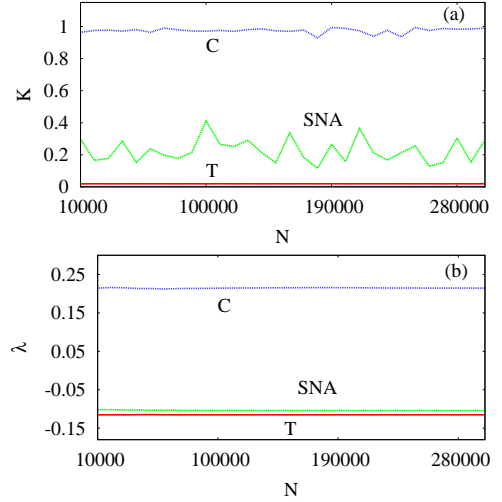}
\caption{\label{fig15}(Color online)Plots of (a) asymptotic growth rate $K$ (b) Lyapunov exponent as a function of time series in the quasiperiodically forced logistic map (Eq.(11)): Here, T indicates torus for $\epsilon'=0.3$, $\alpha=3.4874$; SNA for $\epsilon'=0.3$,  $\alpha=3.488$; chaotic attractor $(C)$ for $\epsilon'=0.5$, $\alpha=3.6$.}
\end{figure}
\end{appendix}

\begin{figure}
\centering
\includegraphics[width=1.00\columnwidth=0.4,height=8cm]{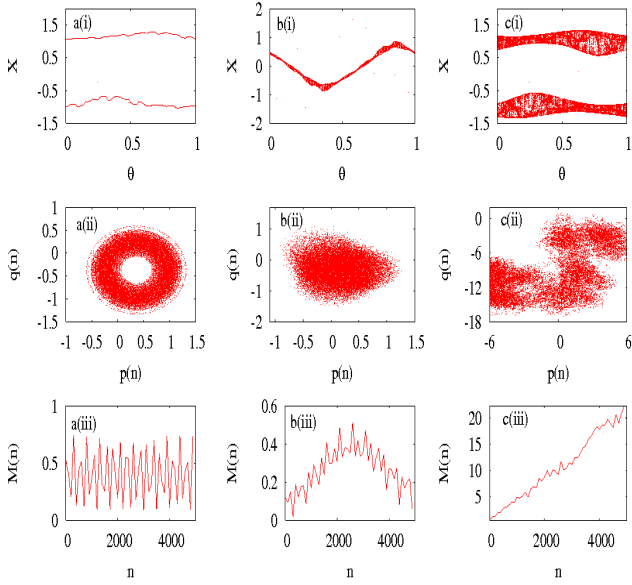}
\caption{\label{fig16}(Color online) For the quasiperiodically forced cubic map (Eq.(12)) (a) Torus for $f=0.10 $, $A=2.14$  (b) SNA for $f=0.7$, $A=1.86687$ and (c) chaotic attractor for $f=0.10$, $A=2.19$: (i) Projection of the attractor; (ii) dynamics of the translation variables $(p,q)$ in terms of 0-1 test (Eqs.(1)-(2)); (iii) Mean square displacement $M(n)$ as a function of n. Here $N=5\times10^{4}$ with $n=N/10$.}
\end{figure}

\begin{appendix}
\section*{Appendix C: Characterization of different attractors of the quasiperiodically forced cubic map using 0-1 test}
In this appendix, we present the characterization of different dynamical attractors in the quasiperiodically forced cubic map (Eq.(12)) in terms of the  translation variables $(p,q)$ and the mean square displacement $M(n)$ using the 0-1 test with $c$ = $\frac{(2\pi)^{2}}{\omega}$, and $\omega=(\sqrt{5}-1)/2$. The results are given in Fig.16. Again this analysis indicates that the 0-1 test allows one to distinguish SNAs from quasiperiodic and chaotic attractors as in the case of the quasiperiodically forced logistic map (Sec. III.B). 
\end{appendix}

\begin{figure}
\centering
\includegraphics[width=0.90\columnwidth=0.7]{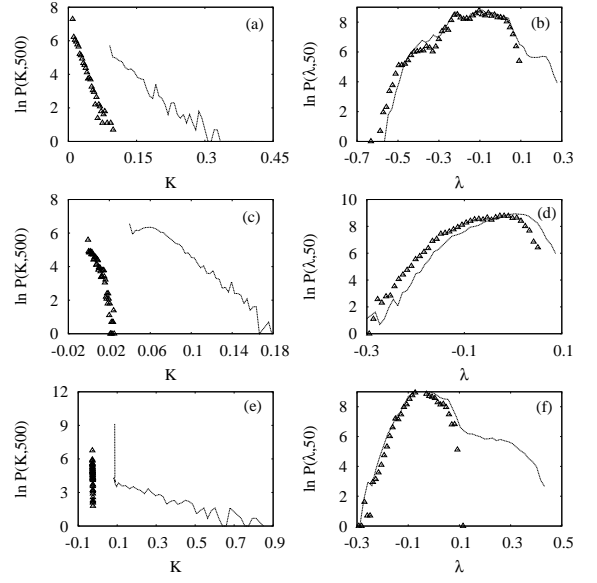}
\caption{\label{fig17}A comparative distribution of asymptotic growth rate $P(K,N)$ as a function of $K$ for the three transitions showing its values for a torus ($\Delta$) just before transition and a SNA (continuous curve) just after transition in the case of the quasiperiodically driven logistic map (Eq. 11): (a) HH route for the parameter $\epsilon'=0.3$, $\alpha=3.4874$ (for torus) and $\alpha=3.488$ (for SNA); (c) Fractalization route $\epsilon'=1.00$, $\alpha=2.63$ (for torus) and $\alpha=2.652$ (SNA); (e) Intermittency route with $\epsilon'=1.00$, and $\alpha=3.40581$ (torus), and $\alpha=3.4058088$ (SNA). The corresponding distribution of $P(\lambda,N)$ as a function of $\lambda$ are given in (b), (d) and (f), respectively, for the above three routes for comparision purpose.}
\end{figure}

\section*{Appendix D: Statistical evaluation of the asymptotic growth rate and finite time Lyapunov exponents(FTLEs)}
In this section we wish to present a statistical evaluation of our results by analyzing the distribution  $P(K,N)$ of local asymptotic growth rate of $K$ and comparing the same with the distribution $P(\lambda,N)$ of FTLEs for each one of the transitions from torus to SNA. The local distribution of $K$ is calculated over short segements of the trajectories of the given dynamical system. Then, the distribution of local asymptotic growth rate of $K$ is defined similar to FTLEs[21] as
\begin{center}
$P(N,K) dK=$Probability that $K$ takes a value between $K$ and $K+dK$.
\end{center}

We have calculated both the distributions of the local asymptotic growth rate $(P(K,N))$ and FTLEs$(P(\lambda,N))$ in order to compare the effectiveness of the two descriptions.  For each one of the routes to SNAs, we have investigated the distributions of local $K$ and FTLES, (i) for a torus prior to the transition to SNAs and (ii) for a SNA just after the transition. We first present the  results for the quasiperiodically forced logistic map in Fig.17. The local distribution of $K$ is given in Figs.17(a), 17(c), and 17(e) and the distribution of FTLEs is given in Figs. 17(b), 17(d) and 17(f) corresponding to HH, fractalization, and intermittency routes, respectively.\\

From Figures 17(a),(c) and (e), one can note that for all the transitions $P(K,N$) tends strongly to zero when the attractor is a torus but on the SNA the distribution spreads between 0.1 and 0.9. In particular, the value of $K$ is distributed from 0.1 to 0.35 for the birth of SNA through HH route. The distribution of $K$ extends from 0.05 to 0.18 for the logistic map (or 0.2 to 0.3 for the cubic map in Fig. 18) for the genesis of SNA through fractalization route. In the intermittent SNA the distribution spreads across the range of $K$ from 0.1 to 0.9. The distribution of different ranges of values of $K$ helps to distinguish the different transitions to SNAs. It is instructive to compare the above results with the distribution of finite time Lyapunov exponents  as shown in Figs.17(b),(d),(f). One can note that the distribution of finite time Lyapunov exponents have the same shape for HH and fractalization route and a slightly different one for the intermittent SNA.  Moreover, the distribution of $P(\lambda,N)$ for the both cases of  torus and SNA are almost identical except for the tails on the positive side of $\lambda$.  However, the distribution of local $K$ clearly distinguishes torus and SNA as well as the three transitions.\\
\hspace*{0.5cm} A similar behavior of the distribution of both local $K$ and FTLEs has been observed for the  cubic map as shown in Fig.18.  For the case of quasiperiodically forced Duffing oscillator also a similar picture arises.
\begin{figure}
\centering
\includegraphics[width=0.9\columnwidth=0.7]{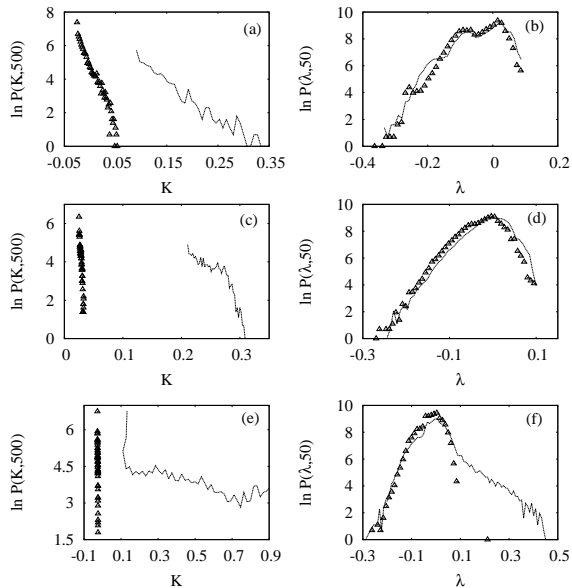}
\caption{\label{fig18} Same as Fig. 17 but now for the case of quasiperiodically forced cubic map (Eq. 12). The choice of parameters here are as follows. (a),(b): HH route with $f=0.7$, and $A=1.8865$ (torus) and $A=1.8870$ (SNA); (c),(d): Fractalization route with $f=0.1$, and $A=2.14$ (torus) and $A=2.167$ (SNA); (e),(f): Intermittency route with $f=0.7$,$A=1.8017$ (torus) and $A=1.801685$ (SNA).}
\end{figure}

\newpage 


\begin{thebibliography}{45}
\expandafter\ifx\csname natexlab\endcsname\relax\def\natexlab#1{#1}\fi
\expandafter\ifx\csname bibnamefont\endcsname\relax
  \def\bibnamefont#1{#1}\fi
\expandafter\ifx\csname bibfnamefont\endcsname\relax
  \def\bibfnamefont#1{#1}\fi
\expandafter\ifx\csname citenamefont\endcsname\relax
  \def\citenamefont#1{#1}\fi
\expandafter\ifx\csname url\endcsname\relax
  \def\url#1{\texttt{#1}}\fi
\expandafter\ifx\csname urlprefix\endcsname\relax\def\urlprefix{URL }\fi
\providecommand{\bibinfo}[2]{#2}
\providecommand{\eprint}[2][]{\url{#2}}

\bibitem[{\citenamefont{Grassberger and Procaccia}(1993)\citenamefont{Grassberger and Procaccia}}]{grassberger1993}
\bibinfo{author}{\bibfnamefont{P.}~\bibnamefont{Grassberger}} \bibnamefont{and}
\bibinfo{author}{\bibfnamefont{I.}~\bibnamefont{Procaccia}}, \bibinfo{journal}{Phys. Rev. Lett} \textbf{\bibinfo{volume}{50}}, \bibinfo{pages}{326} (\bibinfo{year}{1993});
\bibinfo{journal}{Physica D} \textbf{\bibinfo{volume}{9}}, \bibinfo{pages}{189}(\bibinfo{year}{1983}).


\bibitem[{\citenamefont{Ott}(1994)\citenamefont{Ott}}]{ottb1994}
\bibinfo{author}{\bibfnamefont{E.}~\bibnamefont{Ott}}, 
\bibinfo{title}{\emph {Chaos in Dynamical Systems}}, 
(\bibinfo{publisher}{Cambridge University Press, Cambridge}, \bibinfo{address}{England}, \bibinfo{year}{1994}).


\bibitem[{\citenamefont{Lakshmanan et~al}(2006)\citenamefont{Lakshmanan and Rajasekar}}]{lakshmanan2003}
\bibinfo{author}{\bibfnamefont{M.}~\bibnamefont{Lakshmanan}} \bibnamefont{and}
\bibinfo{author}{\bibfnamefont{S.}~\bibnamefont{Rajasekar}}, \bibinfo{title}{\emph{Nonlinear Dynamics:Integrability Chaos and Patterns}}, 
(\bibinfo{publisher}{Springer-verleg}, \bibinfo{address}{Berlin}, \bibinfo{year}{2003}).


\bibitem[{\citenamefont{Gottwald et~al }(2004)\citenamefont{Gottwald, and Melbourne}}]{gottwald2004}
\bibinfo{author}{\bibfnamefont{G.~A.}~\bibnamefont{Gottwald}}, \bibnamefont{and}
 \bibinfo{author}{\bibfnamefont{I.}~\bibnamefont{Melbourne}},
\bibinfo{journale}{Proc. R. Soc. London. Ser. A} \textbf{\bibinfo{volume}{460}},
\bibinfo{pages}{603} (\bibinfo{year}{2004}).



\bibitem[{\citenamefont{Gottwald et~al}(2005)\citenamefont{Gottwald, and Melbourne}}]{gottwald2005}
\bibinfo{author}{\bibfnamefont{G.~A.}~\bibnamefont{Gottwald}}, \bibnamefont{and}
\bibinfo{author}{\bibfnamefont{I.}~\bibnamefont{Melbourne}},
\bibinfo{journal}{Physica D} \textbf{\bibinfo{volume}{212}},
\bibinfo{pages}{110} (\bibinfo{year}{2005}).



\bibitem[{\citenamefont{Gottwald et~al}(2009)\citenamefont{Gottwald and Melbourne}}]{gottwald2009}
\bibinfo{author}{\bibfnamefont{G.~A.}\bibnamefont{Gottwald}} \bibnamefont{and}
\bibinfo{author}{\bibfnamefont{I.}~\bibnamefont{Melbourne}},
\bibinfo{journal}{SIAM J. Appl. Dyn. Syst} \textbf{\bibinfo{volume}{8}},
\bibinfo{pages} (\bibinfo{year}{2009});
\bibinfo{author}{\bibfnamefont{G.~A.}~\bibnamefont{Gottwald}} \bibnamefont{and}
\bibinfo{author}{\bibfnamefont{I.}~\bibnamefont{Melbourne}},
\bibinfo{journal}{Nonlinearity} \textbf{\bibinfo{volume}{22}},
\bibinfo{pages}{1361} (\bibinfo{year}{2009}).



\bibitem[{\citenamefont{Hu et~al}(2005)\citenamefont{Hu, Tung, Gao and Cao}}]{hu2005}
\bibinfo{author}{\bibfnamefont{J.}~\bibnamefont{Hu}},
\bibinfo{author}{\bibfnamefont{W.}~\bibnamefont{Tung}},
\bibinfo{author}{\bibfnamefont{J.}~\bibnamefont{Gao}},  
\bibnamefont{and} {\bibfnamefont{Y.}~\bibnamefont{Cao}}, 
\bibinfo{journal}{Phys. Rev. E} \textbf{\bibinfo{volume}{72}},
\bibinfo{pages}{056207} (\bibinfo{year}{2005}).


\bibitem[{\citenamefont{Christopher and Smith}(2011)\citenamefont{ Christopher and Smith}}]{christopher2011}
\bibinfo{author}{\bibfnamefont{W.~K.}~\bibnamefont{Christopher}} \bibnamefont{and}
\bibinfo{author}{\bibfnamefont{S.}~\bibnamefont{Smith}},
\bibinfo{journal}{Phys. Rev. E} \textbf{\bibinfo{volume}{83}}, 
\bibinfo{pages}{03621} (\bibinfo{year}{2011}).

\bibitem[{\citenamefont{Romera-Bastida and Rayes-Martinez}(2011)\citenamefont{Romera-Bastida and Rayes-Martinez}}]{romera2011}
\bibinfo{author}{\bibfnamefont{M.}~\bibnamefont{Romera-Bastida}} \bibnamefont{and}
\bibinfo{author}{\bibfnamefont{A.~Y.}~\bibnamefont{Rayes-Martinez}},
\bibinfo{journal}{Phys. Rev. E} \textbf{\bibinfo{volume}{83}},
\bibinfo{pages}{016213} (\bibinfo{year}{2011}).


\bibitem[{\citenamefont{Grebogi et~al}(1984)\citenamefont{Grebogi, Ott, Romeiras, and Yorke}}]{grebogi1984}
\bibinfo{author}{\bibfnamefont{C.}~\bibnamefont{Grebogi}},
\bibinfo{author}{\bibfnamefont{E.}~\bibnamefont{Ott}},
\bibinfo{author}{\bibfnamefont{S.}~\bibnamefont{Pelikan}}, \bibnamefont{and}
\bibinfo{author}{\bibfnamefont{J.~A.}\bibnamefont{Yorke}},
\bibinfo{journal}{Physica D } \textbf{\bibinfo{volume}{13}},
\bibinfo{pages}{261} (\bibinfo{year}{1984}); \bibinfo{author}{\bibfnamefont{C.}~\bibnamefont{Grebogi}},
\bibinfo{author}{\bibfnamefont{E.}~\bibnamefont{Ott}},
\bibinfo{author}{\bibfnamefont{F.~J.}~\bibnamefont{Romeiras}}, \bibnamefont{and}
\bibinfo{author}{\bibfnamefont{J.~A.}\bibnamefont{Yorke}},
\bibinfo{journal}{Phys. Rev. A } \textbf{\bibinfo{volume}{36}},
\bibinfo{pages}{5365} (\bibinfo{year}{1987}).

\bibitem[{\citenamefont{Kuznetsov et~al}(1995)\citenamefont{Kuznetsov, Pikovsky, and Feudal}}]{kuznetsov1995}

\bibinfo{author}{\bibfnamefont{S.~P.}~\bibnamefont{Kuznetsov}},
\bibinfo{author}{\bibfnamefont{A.~S.}~\bibnamefont{Pikovsky}},
\bibnamefont{and} \bibinfo{author}{\bibfnamefont{U.}~\bibnamefont{Feudel}},
\bibinfo{journal}{Phys. Rev. E } \textbf{\bibinfo{volume}{51}},
\bibinfo{pages}{R1629} (\bibinfo{year}{1995}).

\bibitem[{\citenamefont{Romeiras and Ott}(1987)\citenamefont{Romeiras and Ott}}]{romeiras1987}
\bibinfo{author}{\bibfnamefont{F.~J}~\bibnamefont{Romeiras}},
\bibinfo{author}{\bibfnamefont{E.}~\bibnamefont{Ott}},
\bibinfo{journal}{Phys. Rev. A} \textbf{\bibinfo{volume}{35}},
\bibinfo{pages}{4404}, (\bibinfo{year}{1987}).

\bibitem[{\citenamefont{Ding et~al}(1989)\citenamefont{Ding, Grebogi and Ott}}]{ding1989}
\bibinfo{author}{\bibfnamefont{M.}~\bibnamefont{Ding}},
\bibinfo{author}{\bibfnamefont{C.}~\bibnamefont{Grebogi}}, \bibnamefont{and}
\bibinfo{author}{\bibfnamefont{E.}~\bibnamefont{Ott}},
\bibinfo{journal}{Phys. Rev. A} \textbf{\bibinfo{volume}{39}},
\bibinfo{pages}{2593} (\bibinfo{year}{1989}).

\bibitem[{\citenamefont{Bondeson et~al}(1985)\citenamefont{Bondeson, Ott and Antonsen}}]{bondeson1985}
\bibinfo{author}{\bibfnamefont{A.}~\bibnamefont{Bondeson}},
\bibinfo{author}{\bibfnamefont{E.}~\bibnamefont{Ott}}, \bibnamefont{and}
\bibinfo{author}{\bibfnamefont{T.~M.}\bibnamefont{Antonsen}},
\bibinfo{journal}{Phys. Rev. Lett.} \textbf{\bibinfo{volume}{55}},
\bibinfo{pages}{2103} (\bibinfo{year} {1985}).

\bibitem[{\citenamefont{Ketoja and Satija}(1997)\citenamefont{Ketoja and Satija}}]{ketoja1997}
\bibinfo{author}{\bibfnamefont{J.~A.} \bibnamefont{Ketoja}} \bibnamefont{and}
\bibinfo{author}{\bibfnamefont{I.}~\bibnamefont{Satija}},
\bibinfo{journal}{Physica D} \textbf{\bibinfo{volume}{109}},
\bibinfo{pages}{70} (\bibinfo{year}{1997}).

\bibitem[{\citenamefont{Ditto et~al}(1990)\citenamefont{Ditto, Spano, Savage, Rauseo, Heagy, and Ott}}]{ditto1990}
\bibinfo{author}{\bibfnamefont{W.~L.}~\bibnamefont{Ditto}},
\bibinfo{author}{\bibfnamefont{M.~L.}~\bibnamefont{Spano}},
\bibinfo{author}{\bibfnamefont{H.~T.}~\bibnamefont{Savage}},
\bibinfo{author}{\bibfnamefont{S.~N.}~\bibnamefont{Rauseo}},
\bibinfo{author}{\bibfnamefont{J.}~\bibnamefont{Heagy}}, \bibnamefont{and}
\bibinfo{author}{\bibfnamefont{E.}~\bibnamefont{Ott}},
\bibinfo{journal}{Phy. Rev. Lett.} \textbf{\bibinfo{volume}{65}},
\bibinfo{pages}{533} (\bibinfo{year}{1990}).

\bibitem[{\citenamefont{Zhou et~al}(1992)\citenamefont{Zhou, Moss, and Bulsara}}]{zhou1992}

\bibinfo{author}{\bibfnamefont{T.}~\bibnamefont{Zhou}},
\bibinfo{author}{\bibfnamefont{F.}~\bibnamefont{Moss}}, \bibnamefont{and}
\bibinfo{author}{\bibfnamefont{A.}~\bibnamefont{Bulsara}},
\bibinfo{journal}{Phys. Rev. A} \textbf{\bibinfo{volume}{45}},
\bibinfo{pages}{5394} (\bibinfo{year}{1992}).

\bibitem[{\citenamefont{Ding et~al}(1997)\citenamefont{Ding, Deutsch, Dinklage, and Wilke  }}]{ding1997}

\bibinfo{author}{\bibfnamefont{W.~X} \bibnamefont{Ding}},
\bibinfo{author}{\bibfnamefont{H.}~\bibnamefont{Deutsch}},
\bibinfo{author}{\bibfnamefont{A.}~\bibnamefont{Dinklage}}, \bibnamefont{and}
\bibinfo{author}{\bibfnamefont{C.}~\bibnamefont{Wilke}},
\bibinfo{journal}{Phys. Rev. E} \textbf{\bibinfo{volume}{55}},
\bibinfo{pages}{3769} (\bibinfo{year}{1997}).

\bibitem[{\citenamefont{Ruiz and Parmananda}(2007)\citenamefont{Ruiz and Parmananda}}]{ruiz2007}

\bibinfo{author}{\bibfnamefont{G.}~\bibnamefont{Ruiz}} \bibnamefont{and}
\bibinfo{author}{\bibfnamefont{P.}~\bibnamefont{Parmananda}},
\bibinfo{journal}{Phys. Lett. A} \textbf{\bibinfo{volume}{367}},
\bibinfo{pages}{478} (\bibinfo{year}{2007}).

\bibitem[{\citenamefont{Thamilmaran et~al}(2006)\citenamefont{Thamilmaran, Senthilkumar, Venkatesan, and Lakshmanan}}]{thamilmaran2006}

\bibinfo{author}{\bibfnamefont{K.}~ \bibnamefont{Thamilmaran}},
\bibinfo{author}{\bibfnamefont{D.~V.} \bibnamefont{Senthilkumar}},
\bibinfo{author}{\bibfnamefont{A.}~\bibnamefont{Venkatesan}}, \bibnamefont{and}
\bibinfo{author}{\bibfnamefont{M.}~\bibnamefont{Lakshmanan}},
\bibinfo{journal}{Phys. Rev. E} \textbf{\bibinfo{volume}{74}},
\bibinfo{pages}{036205}, (\bibinfo{year}{2006});
\bibinfo{author}{\bibfnamefont{D.~V.} \bibnamefont{Senthilkumar}},
\bibinfo{author}{\bibfnamefont{K.}~\bibnamefont{Srinivasan}},
\bibinfo{author}{\bibfnamefont{K.}~\bibnamefont{Thamilmaran}},\bibnamefont{and}
\bibinfo{author}{\bibfnamefont{M.}~\bibnamefont{Lakshmanan}},
\bibinfo{journal}{Phys. Rev. E} \textbf{\bibinfo{volume}{78}},
\bibinfo{pages}{066211} (\bibinfo{year}{2008}).



\bibitem[{\citenamefont{Yang and Bilimgut}(1997)\citenamefont{Yang and Bilimgut}}]{yang1997}

\bibinfo{author}{\bibfnamefont{T.}~\bibnamefont{Yang}}, \bibnamefont{and}
\bibinfo{author}{\bibfnamefont{K.}~\bibnamefont{Bilimgut}},
\bibinfo{journal}{Phys. Lett. A} \textbf{\bibinfo{volume}{236}},
\bibinfo{pages}{494}, (\bibinfo{year}{1997}); 
 \bibinfo{author}{\bibfnamefont{Z.}~\bibnamefont{Liu}}  \bibnamefont{and}
\bibinfo{author}{\bibfnamefont{Z.}~\bibnamefont{Zhua}},
\bibinfo{journal}{Int. J. Bifurcations Chaos Appl. Sci. Eng} \textbf{\bibinfo{volume}{6}},
\bibinfo{pages}{1383} (\bibinfo{year}{1996});

\bibitem[{\citenamefont{Prasad et~al}(1998)\citenamefont{Prasad, Mehra, and Ramasamy}}]{prasad1998}
\bibinfo{author}{\bibfnamefont{A.}~\bibnamefont{Prasad}},
\bibinfo{author}{\bibfnamefont{V.}~\bibnamefont{Mehra}}, \bibnamefont{and}
\bibinfo{author}{\bibfnamefont{R.}~\bibnamefont{Ramasamy}},
\bibinfo{journal}{Phys. Rev. E } \textbf{\bibinfo{volume}{57}},
\bibinfo{pages}{1156}, (\bibinfo{year}{1998}).
\bibinfo{author}{\bibfnamefont{A.}~\bibnamefont{Prasad}},
\bibinfo{author}{\bibfnamefont{V.}~\bibnamefont{Mehra}}, \bibnamefont{and}
\bibinfo{author}{\bibfnamefont{R.}~\bibnamefont{Ramasamy}},
\bibinfo{journal}{Phys. Rev. Lett } \textbf{\bibinfo{volume}{79}},
\bibinfo{pages}{4127}  (\bibinfo{year}{1997}).

\bibitem[{\citenamefont{Venkatesan et~al}(2000)\citenamefont{Venkatesan, Lakshmanan, Prasad, and Ramasamy}}]{venkatesan2000}
\bibinfo{author}{\bibfnamefont{A.}~\bibnamefont{Venkatesan}},
\bibinfo{author}{\bibfnamefont{M.}~\bibnamefont{Lakshmanan}},
\bibinfo{author}{\bibfnamefont{A.}~\bibnamefont{Prasad}}, \bibnamefont{and}
\bibinfo{author}{\bibfnamefont{R.}~\bibnamefont{Ramaswamy}},
\bibinfo{journal}{Phys. Rev. E } \textbf{\bibinfo{volume}{61}},
\bibinfo{pages}{3641}  (\bibinfo{year}{2000}).


\bibitem[{\citenamefont{Heagy and Hammel}(1994)\citenamefont{Heagy and Hammel}}]{heagy1994}

\bibinfo{author}{\bibfnamefont{J.~F.} \bibnamefont{Heagy}}, \bibnamefont{and}
\bibinfo{author}{\bibfnamefont{H.}~\bibnamefont{Hammel}},
\bibinfo{journal}{Physica D } \textbf{\bibinfo{volume}{70}},
\bibinfo{pages}{140} (\bibinfo{year}{1994}).

\bibitem[{\citenamefont{Ott and Sommerer}(1994)\citenamefont{Ott and  Sommerer}}]{ott1994}
\bibinfo{author}{\bibfnamefont{E.}~\bibnamefont{Ott}} \bibnamefont{and}
\bibinfo{author}{\bibfnamefont{J.~C.}~\bibnamefont{Sommerer}},
\bibinfo{journal}{Phys. Rev. Lett A } \textbf{\bibinfo{volume}{188}},
\bibinfo{pages}{39} (\bibinfo{year}{1994}).

\bibitem[{\citenamefont{Kaneko}(1984)\citenamefont{Kaneko}}]{kaneko1984}
\bibinfo{author}{\bibfnamefont{K.}~\bibnamefont{Kaneko}},
\bibinfo{journal}{Theor. Phys } \textbf{\bibinfo{volume}{71}},
\bibinfo{pages}{1112} (\bibinfo{year}{1984});
\bibinfo{journal}{Prog. Theor. Phys } \textbf{\bibinfo{volume}{71}},
\bibinfo{pages}{140} (\bibinfo{year}{1994});



\bibitem[{\citenamefont{Nishikawa, Kaneko}(1996)\citenamefont{Nishikawa, Kaneko}}]{nishikawa1996}
\bibinfo{author}{\bibfnamefont{T.}~\bibnamefont{Nishikawa}},
\bibinfo{author}{\bibfnamefont{K.}~\bibnamefont{Kaneko}},
\bibinfo{journal}{Phys. Rev. E} \textbf{\bibinfo{volume}{54}},
\bibinfo{pages}{1063} (\bibinfo{year}{1996});
\bibinfo{journal}{Phys. Rev. E} \textbf{\bibinfo{volume}{54}},
\bibinfo{pages}{6114} (\bibinfo{year}{1996})

\bibitem[{\citenamefont{Venkatesan and Lakshmanan}(1998)\citenamefont{Venkatesan and  Lakshmanan}}]{venkatesan1998}
\bibinfo{author}{\bibfnamefont{A.}~\bibnamefont{Venkatesan}} \bibnamefont{and}
\bibinfo{author}{\bibfnamefont{M.}~\bibnamefont{Lakshmanan}},
\bibinfo{journal}{Phys. Rev. E } \textbf{\bibinfo{volume}{58}},
\bibinfo{pages}{3008} (\bibinfo{year}{1998}).



\bibitem[{\citenamefont{Venkatesan et~al}(1999)\citenamefont{Venkatesan, Murali, and Lakshmanan}}]{venkatesan1999}

\bibinfo{author}{\bibfnamefont{A.}~\bibnamefont{Venkatesan}},
\bibinfo{author}{\bibfnamefont{K.}~\bibnamefont{Murali}} \bibnamefont{and}
\bibinfo{author}{\bibfnamefont{M.}~\bibnamefont{Lakshmanan}},
\bibinfo{journal}{Phys. Lett A } \textbf{\bibinfo{volume}{259}},
\bibinfo{pages}{246} (\bibinfo{year}{1999}).

\bibitem[{\citenamefont{Ngamga et~al}(2007)\citenamefont{Ngamga, Nandi, Ramaswamy, Romano, Thiel, and Kurths}}]{ngamga2007}
\bibinfo{author}{\bibfnamefont{E.~J.} \bibnamefont{Ngamga}},
\bibinfo{author}{\bibfnamefont{A.}~\bibnamefont{Nandi}},
\bibinfo{author}{\bibfnamefont{R.}~\bibnamefont{Ramaswamy}},
\bibinfo{author}{\bibfnamefont{M.~C.} \bibnamefont{Romano}},
\bibinfo{author}{\bibfnamefont{M.}~\bibnamefont{Thiel}}, \bibnamefont{and}
\bibinfo{author}{\bibfnamefont{J.}~\bibnamefont{Kurths}},
\bibinfo{journal}{Phys. Rev. E } \textbf{\bibinfo{volume}{75}},
\bibinfo{pages}{036222} (\bibinfo{year}{2007});
\bibinfo{author}{\bibfnamefont{E.~J.} \bibnamefont{Ngamga}},
\bibinfo{author}{\bibfnamefont{D.~V.} \bibnamefont{Senthilkumar}}, \bibnamefont{and} 
\bibinfo{author}{\bibfnamefont{J.}~\bibnamefont{Kurths}},
\bibinfo{journal}{Eur. Phys. J. Spec. Top} \textbf{\bibinfo{volume}{191}},
\bibinfo{pages}{15} (\bibinfo{year}{2011});
\bibinfo{author}{\bibfnamefont{E.~J.} \bibnamefont{Ngamga}},
\bibinfo{author}{\bibfnamefont{D.~V.} \bibnamefont{Senthilkumar}}, 
\bibinfo{author}{\bibfnamefont{A.}~\bibnamefont{Prasad}},
\bibinfo{author}{\bibfnamefont{P.}~\bibnamefont{Paramananda}},
\bibinfo{author}{\bibfnamefont{N.}~\bibnamefont{Marvan}}, \bibnamefont{and} 
\bibinfo{author}{\bibfnamefont{J.}~\bibnamefont{Kurths}},
\bibinfo{journal}{Phys. Rev. E} \textbf{\bibinfo{volume}{85}},
\bibinfo{pages}{026217} (\bibinfo{year}{2012});


\bibitem[{\citenamefont{Venkatesan and Lakshmanan}(2001)\citenamefont{Venkatesan and  Lakshmanan}}]{venkatesan2001}
\bibinfo{author}{\bibfnamefont{A.}~\bibnamefont{Venkatesan}} \bibnamefont{and}
\bibinfo{author}{\bibfnamefont{M.}~\bibnamefont{Lakshmanan}}, 
\bibinfo{journal}{Phys. Rev. E } \textbf{\bibinfo{volume}{63}},
\bibinfo{pages}{026219} (\bibinfo{year}{2001}).

\bibitem[{\citenamefont{Feudel et~al}(2006)\citenamefont{Feudel, Kuznetsov, and Pikovsky}}]{feudel2006}
\bibinfo{author}{\bibfnamefont{U.}~\bibnamefont{Feudel}},
\bibinfo{author}{\bibfnamefont{S.}~\bibnamefont{Kuznetsov}}, \bibnamefont{and} 
\bibinfo{author}{\bibfnamefont{A.}~\bibnamefont{Pikovsky}},
 \bibinfo{title}{\emph{Strange Nonchaotic Attractors: Dynamics between Order and Chaos in Quasiperiodically Forced Systems}}, (\bibinfo{publisher}{Worls Scientific},\bibinfo{address}{Singapore}, \bibinfo{year}{2006}).


\bibitem[{\citenamefont{Prasad et~al}(2001)\citenamefont{Prasad,Negi and Ramaswamy}}]{prasad2001}

\bibinfo{author}{\bibfnamefont{A.}~\bibnamefont{Prasad}},
\bibinfo{author}{\bibfnamefont{S~.S}~\bibnamefont{Negi}}, \bibnamefont{and} 
\bibinfo{author}{\bibfnamefont{R.}~\bibnamefont{Ramaswamy}}, 
\bibinfo{journal}{Int. J. Bifurc. Chaos. Appl. Sci. Eng} \textbf{\bibinfo{volume}{11}},
\bibinfo{pages}{291} (\bibinfo{year}{2001}).


\bibitem[{\citenamefont{Kim et~al}(2003)\citenamefont{Kim, Lim and Ott}}]{kim2003}

\bibinfo{author}{\bibfnamefont{S.~Y.}~\bibnamefont{Kim}},
\bibinfo{author}{\bibfnamefont{W.}~\bibnamefont{Lim}}, \bibnamefont{and} 
\bibinfo{author}{\bibfnamefont{E.}~\bibnamefont{Ott}}, 
\bibinfo{journal}{Phys. Rev. E} \textbf{\bibinfo{volume}{63}},
\bibinfo{pages}{056203} (\bibinfo{year}{2003}).



\bibitem[{\citenamefont{Lai}(1996)\citenamefont{Lai}}]{lai1996}

\bibinfo{author}{\bibfnamefont{Y.~C.} \bibnamefont{Lai}}, 
\bibinfo{journal}{Phys. Rev. E} \textbf{\bibinfo{volume}{53}},
\bibinfo{pages}{57} (\bibinfo{year}{1996}).

\bibitem[{\citenamefont{Yalcinkaya and Lai}(1998)\citenamefont{Yalcinkaya and Lai}}]{yalcinkaya1998}

\bibinfo{author}{\bibfnamefont{T.}~\bibnamefont{Yalcinkaya}} \bibnamefont{and}
\bibinfo{author}{\bibfnamefont{Y.~C.} \bibnamefont{Lai}},
\bibinfo{journal}{Phys. Rev. Lett } \textbf{\bibinfo{volume}{77}},
\bibinfo{pages}{5039} (\bibinfo{year}{1996}).

\bibitem[{\citenamefont{Wang et~al}(2004)\citenamefont{Wang, Zhan, Lai, and Lai}}]{wang2004}
\bibinfo{author}{\bibfnamefont{X.}~\bibnamefont{Wang}},
{\bibfnamefont{M.}~\bibnamefont{Zhan}},
{\bibfnamefont{C.~H.}~\bibnamefont{Lai}}, \bibnamefont{and}
\bibinfo{author}{\bibfnamefont{Y.~C.} \bibnamefont{Lai}},
\bibinfo{journal}{Phys. Rev. Lett} \textbf{\bibinfo{volume}{92}},
\bibinfo{pages}{074102}, (\bibinfo{year}{2004});
\bibinfo{author}{\bibfnamefont{X.}~\bibnamefont{Wang}},
{\bibfnamefont{Y.C}~\bibnamefont{Lai}}, \bibnamefont{and}
\bibinfo{author}{\bibfnamefont{C.~H} \bibnamefont{Lai}},
\bibinfo{journal}{Phys. Rev. E} \textbf{\bibinfo{volume}{74}},
\bibinfo{pages}{016203} (\bibinfo{year}{2006}).

\bibitem[{\citenamefont{Khavanov et~al}(2000)\citenamefont{Khavanov, Khavanova, Mcchintock, and Anischenko}}]{khavanov2000}
\bibinfo{author}{\bibfnamefont{A.}~ \bibnamefont{Khavanov}},
{\bibfnamefont{N.~A} \bibnamefont{Khavanova}},
{\bibfnamefont{P.~V.~E} \bibnamefont{Mcchintock}}, \bibnamefont{and}
\bibinfo{author} {\bibfnamefont{S.}~\bibnamefont{Anischenko}},
\bibinfo{journal}{Phys. Lett. A} \textbf{\bibinfo{volume}{268}},
\bibinfo{pages}{} (\bibinfo{year}{2000}).

\bibitem[{\citenamefont{Feudel et~al}(1995)\citenamefont{Feudel, Kurths,  and Pikovsky}}]{feudel1995}
\bibinfo{author}{\bibfnamefont{U.}~\bibnamefont{Feudel}},
\bibinfo{author}{\bibfnamefont{J.}~\bibnamefont{Kurths}}, \bibnamefont{and}
\bibinfo{author}{\bibfnamefont{A.}~\bibnamefont{Pikovsky}},
\bibinfo{journal}{Physica D} \textbf{\bibinfo{volume}{88}},
\bibinfo{pages}{176}, (\bibinfo{year}{1995});
\bibinfo{author}{\bibfnamefont{U.}~\bibnamefont{Feudel}},
\bibinfo{author}{\bibfnamefont{C.}~\bibnamefont{Greboji}}, \bibnamefont{and}
\bibinfo{author}{\bibfnamefont{E.}~\bibnamefont{Ott}},
\bibinfo{journal}{Phys. Rep} \textbf{\bibinfo{volume}{290}},
\bibinfo{pages}{11},(\bibinfo{year}{1997}).

\bibitem[{\citenamefont{Pikovsky and Feudel}(1995)\citenamefont{Pikovsky and Feudel}}]{pikovsky1995}
\bibinfo{author}{\bibfnamefont{A.}~\bibnamefont{Pikovsky}} \bibnamefont{and}
\bibinfo{author}{\bibfnamefont{U.}~\bibnamefont{Feudel}},
\bibinfo{journal}{Chaos} \textbf{\bibinfo{volume}{5}},
\bibinfo{pages}{253} (\bibinfo{year}{1995}).

\bibitem[{\citenamefont{Negi et al}(2000)\citenamefont{Negi,Prasad and Ramaswamy}}]{negi2000}
\bibinfo{author}{\bibfnamefont{S~.S.} \bibnamefont{Negi}},
\bibinfo{author}{\bibfnamefont{A.}~\bibnamefont{Prasad}}, \bibnamefont{and} 
\bibinfo{author}{\bibfnamefont{R.}~\bibnamefont{Ramaswamy}}
\bibinfo{journal}{Physica D} \textbf{\bibinfo{volume}{145}},
\bibinfo{pages}{1} (\bibinfo{year}{2000}).
\bibitem[{\citenamefont{Nicol et~al}(2001)\citenamefont{Nicol, Melbourne and Ashwin}}]{nicol2001}
\bibinfo{author}{\bibfnamefont{M.}~\bibnamefont{Nicol}},
\bibinfo{author}{\bibfnamefont{I.}~\bibnamefont{Melbourne}},\bibnamefont{and}{
\bibfnamefont{P.}~\bibnamefont{Ashwin}}, \bibinfo{journal}{Nonlinearity} \textbf{\bibinfo{volume}{14}},
\bibinfo{pages}{275} (\bibinfo{year}{2001}).



\bibitem[{\citenamefont{Ashwin et~al}(2001)\citenamefont{Ashwin, Melbourne and Nicol}}]{ashwin2001}
\bibinfo{author}{\bibfnamefont{P.}~\bibnamefont{Ashwin}},
\bibinfo{author}{\bibfnamefont{I.}~\bibnamefont{Melbourne}}, \bibnamefont{and}
\bibinfo{author}{\bibfnamefont{M.}~\bibnamefont{Nicol}}, \bibinfo{journal}{Physica D} 
\textbf{\bibinfo{volume}{14}} \bibinfo{pages}{275} (\bibinfo{year}{2001}).



\bibitem[{\citenamefont{Dawes et~al}(2008)\citenamefont{Dawes and Freeland}}]{dawes2008}
\bibinfo{author}{\bibfnamefont{J.~H.~P.}~\bibnamefont{Dawes}} \bibnamefont{and} 
\bibinfo{author}{\bibfnamefont{M.~C.}~\bibnamefont{Freeland}}, 
\bibinfo{title}{The 0-1 test for chaos
 and strange nonchaotic attractors};
 \bibinfo{journal}{people.bath.ac.uk/jhpd20/publications} (\bibinfo{year}{2008}); 

\bibitem[{\citenamefont{Press}(1992)}]{press1992}
\bibinfo{author}{\bibfnamefont{W.~H.}~\bibnamefont{Press}},
\bibinfo{author}{\bibfnamefont{S.~A.}~\bibnamefont{Teukolsky}},
\bibinfo{author}{\bibfnamefont{W.~T.}~\bibnamefont{Vetterling}}, \bibnamefont{and}
\bibinfo{author}{\bibfnamefont{B.~P.}~\bibnamefont{Flannery}},
\bibinfo{title}{\emph{Numerical Recipes in C }}
(\bibinfo{publisher}{Cambridge University Press}, \bibinfo{address}{Cambridge}, \bibinfo{year}{2002}).



\end{thebibliography}
\end{document}